\documentclass[aps,prx,superscriptaddress,twocolumn,amsmath,amssymb,longbibliography]{revtex4-2}
\usepackage[T1]{fontenc}
\usepackage[utf8]{inputenc}
\usepackage{babel}
\usepackage{amsmath}
\usepackage{amssymb}
\usepackage{graphicx}
\usepackage[pdfusetitle,colorlinks=true,allcolors=blue]{hyperref}
\usepackage{pifont}

\usepackage{xcolor}

\begin{document}

\title{Statistical mechanics of episodic memory in LLMs}

\author{Huck Stepanyants}
\affiliation{Department of Physics, Harvard University, Cambridge, MA 02138}

\author{Xu Pan}
\affiliation{Center for Brain Science, Harvard University, Cambridge, MA 02138}

\author{Daniel Lee}
\affiliation{Department of Electrical and Computer Engineering, Cornell Tech, New York, NY 10044}

\author{Haim Sompolinsky}
\email{hsompolinsky@mcb.harvard.edu}
\affiliation{Center for Brain Science, Harvard University, Cambridge, MA 02138}
\affiliation{Edmond and Lily Safra Center for Brain Sciences, Hebrew University, Jerusalem 9190401, Israel}

\begin{abstract}
Episodic memories unfold as ordered sequences of correlated events that must be stored in a network which already embodies extensive knowledge of the world, a setting that classical associative memory theories do not address. The brain is thought to employ complementary learning systems for this task: each new episode is stored rapidly in the hippocampus as its own engram, on top of slowly accumulated cortical knowledge. In such a system, the synaptic cost of storing each episode limits how many can be stored over a lifetime, making efficient storage essential. Recently, it was demonstrated that large language models (LLMs) can efficiently store episodic memories in a continual manner by assigning each episode its own gated low-rank adaptation (LoRA) module. Yet why this method is so efficient, and how the storage cost depends on network size, representation dimension, and the number of possible events, remain unclear. To address these questions, we develop a theory of episodic memory storage by LoRA fine-tuning of an LLM's penultimate weight matrix and solve it using geometric arguments and statistical mechanics. We show that the nonconvexity of the storage constraints leads to replica symmetry breaking in the space of connection weights, and that the one-step replica symmetry breaking solution agrees closely with simulations. We further show that LoRA can store sequences of length~${P}$ with~${O(1)}$ parameters per association rather than the~${O(P)}$ of classical recurrent models, and that it naturally supports a large output vocabulary, with capacity decreasing only logarithmically in the vocabulary size. Finally, we demonstrate that the pretrained LLM weights convert a retrieval margin that would otherwise vanish into a finite one. A solvable model of pretrained relative biases explains this effect and reveals an emergent form of prediction-error coding wherein each association’s representation strength grows with its surprise under the pretrained network.
\end{abstract}

\maketitle

\section{Introduction}

Understanding how neural networks store and recall long, correlated sequences is a fundamental problem in both neuroscience and artificial intelligence. Classical recurrent theories of associative memory provide a quantitative framework for storing and retrieving collections of patterns, but most consider relatively simple, uncorrelated memories held in dedicated neural circuits~\citep{hopfield1982,amit1985}. Extensions of classical models to analog, correlated memories and temporal sequences remain an active area of research~\citep{sompolinsky1986,schonsberg2021,cohen2025}. Episodic memories pose a different challenge: each binds familiar concepts into a new combination tied to a specific spatial and temporal context, often in the form of an ordered series of events, and must be incorporated into a network that already represents those concepts~\citep{tulving2002,eichenbaum2004}.

In the brain, this problem is thought to be solved by complementary learning systems~\citep{mcclelland1995}: the hippocampus rapidly encodes new episodes on top of structured knowledge slowly accumulated in the neocortex. According to hippocampal indexing theory, the hippocampus does not store the content of an experience, but rather a compact index into the cortical representations activated by the experience, which it reinstates during recall~\citep{teyler1986}. The population of cells that holds a given memory constitutes its \emph{engram}~\citep{josselyn2020}. But if each episode is stored in its own engram, then the number of new synapses one episode requires limits how many episodes can be stored over a lifetime, making efficient storage of individual episodes essential. How the brain achieves such efficient storage by modifying an existing network, rather than constructing a new memory circuit from scratch, remains poorly understood.

Pretrained large language models (LLMs) offer a concrete setting in which to study this problem. They can memorize and reproduce long sequences of text, demonstrating a remarkable capacity for sequential memory~\citep{carlini2021}. However, much like biological neural systems, the complexity of LLM architectures and their pretraining procedures makes it difficult to identify the minimal mechanisms responsible for this capability. Here, we consider a pretrained LLM as a model system for episodic memory, in which the events of an episode are represented by tokens and new memories are acquired by fine-tuning an existing computational substrate. The pretrained model provides three key ingredients: contextual representations of incoming event sequences through attention, a built-in sequence generator through next-token prediction, and prior knowledge of the semantic content of the events, which can facilitate their encoding and retrieval.

\begin{figure*}[t]
  \centering
  \includegraphics[width=0.9\textwidth]{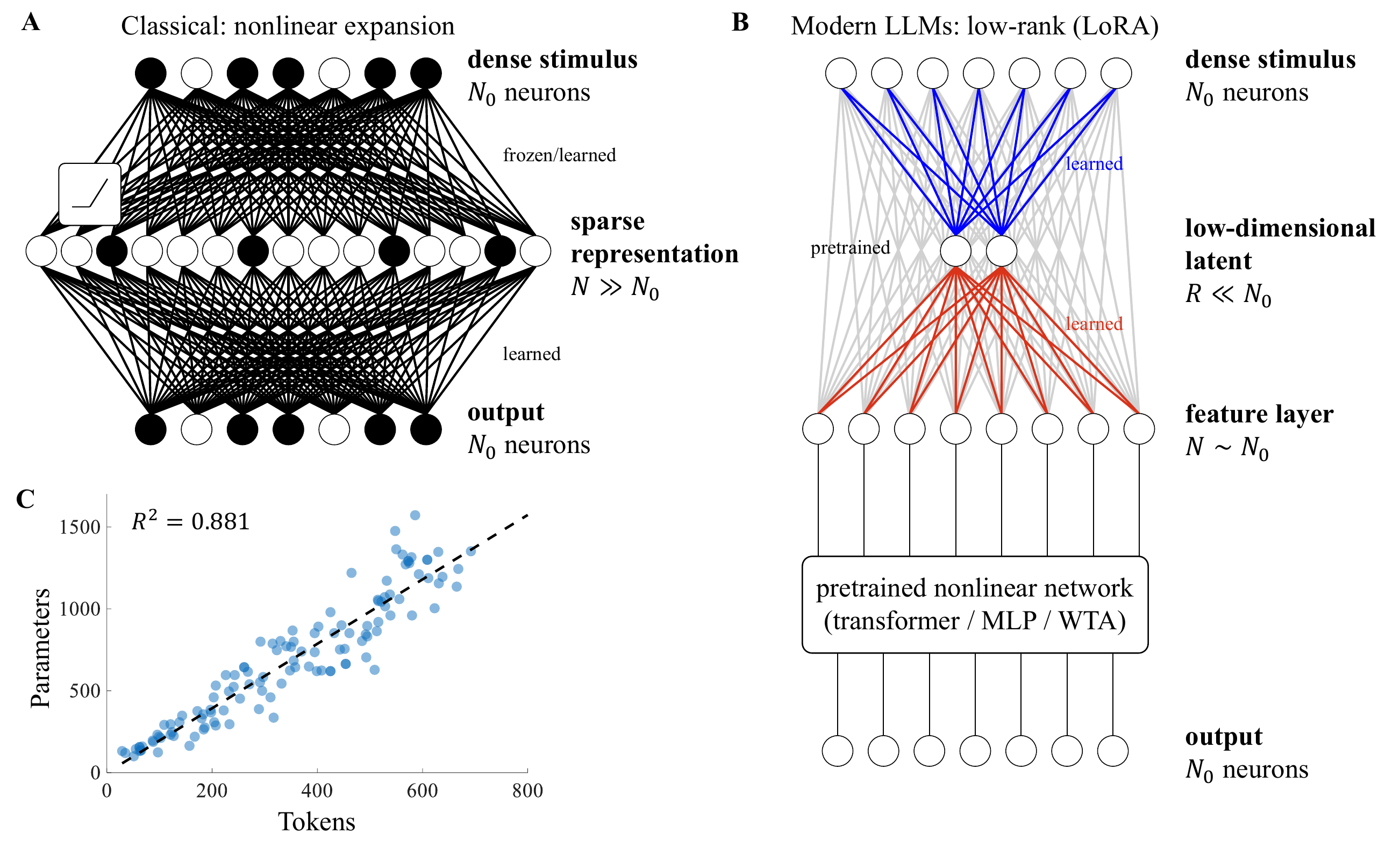}\caption{\textbf{A.} Classical, brain-inspired computational architecture from~\citep{babadi2014}. \textbf{B.} General LoRA architecture. For simplicity, both architectures are considered here for a sequence generation task: the output represents the next event in the same~$N_0$-dimensional space as the dense input layer and can be fed back as input. In the LLM case, this output is the embedding of the predicted next token. \textbf{C.} The minimal number of trainable parameters needed to memorize a sequence with LoRA scales linearly with sequence length (model: Llama-3.1-8B-Instruct). Adapted from Fig.~4B in \citep{pan2026}.}\label{fig:1}
\end{figure*}

We store new memories using low-rank adaptation (LoRA), the most widely used method for fine-tuning pretrained LLMs with few trainable parameters, in which plastic low-rank weight matrices are added to the frozen pretrained weights~\citep{hu2022lora}. In analogy to the brain, the pretrained network plays the role of the neocortex, and each low-rank adapter that of a hippocampal engram. LoRA takes the opposite approach to classical feedforward memory models, which expand the input into a sparse, high-dimensional representation, an architecture found in many sensory processing systems in the brain~\citep{babadi2014,litwinkumar2017} (Fig.~\ref{fig:1}A). Instead, LoRA compresses the input through a low-rank bottleneck (Fig.~\ref{fig:1}B), minimizing the number of modifiable parameters. This economy naturally supports continual learning of episodes: recent work has shown that assigning each episode its own adapter, together with a gating circuit that activates the adapter most relevant to an incoming retrieval cue, allows LLMs to store a long stream of episodes while minimizing catastrophic forgetting~\citep{pan2026}. As in the brain, the number of connections required to store individual episodic memories limits how many can be stored in total. Yet it is not obvious how such compressed representations can store long sequences. Nevertheless, empirical evidence suggests that the number of trainable parameters needed to memorize a sequence with LoRA grows only linearly with its length, even at rank as small as~$2$ (Fig.~\ref{fig:1}C).

\begin{figure*}[t]
  \centering
  \includegraphics[width=0.9\textwidth]{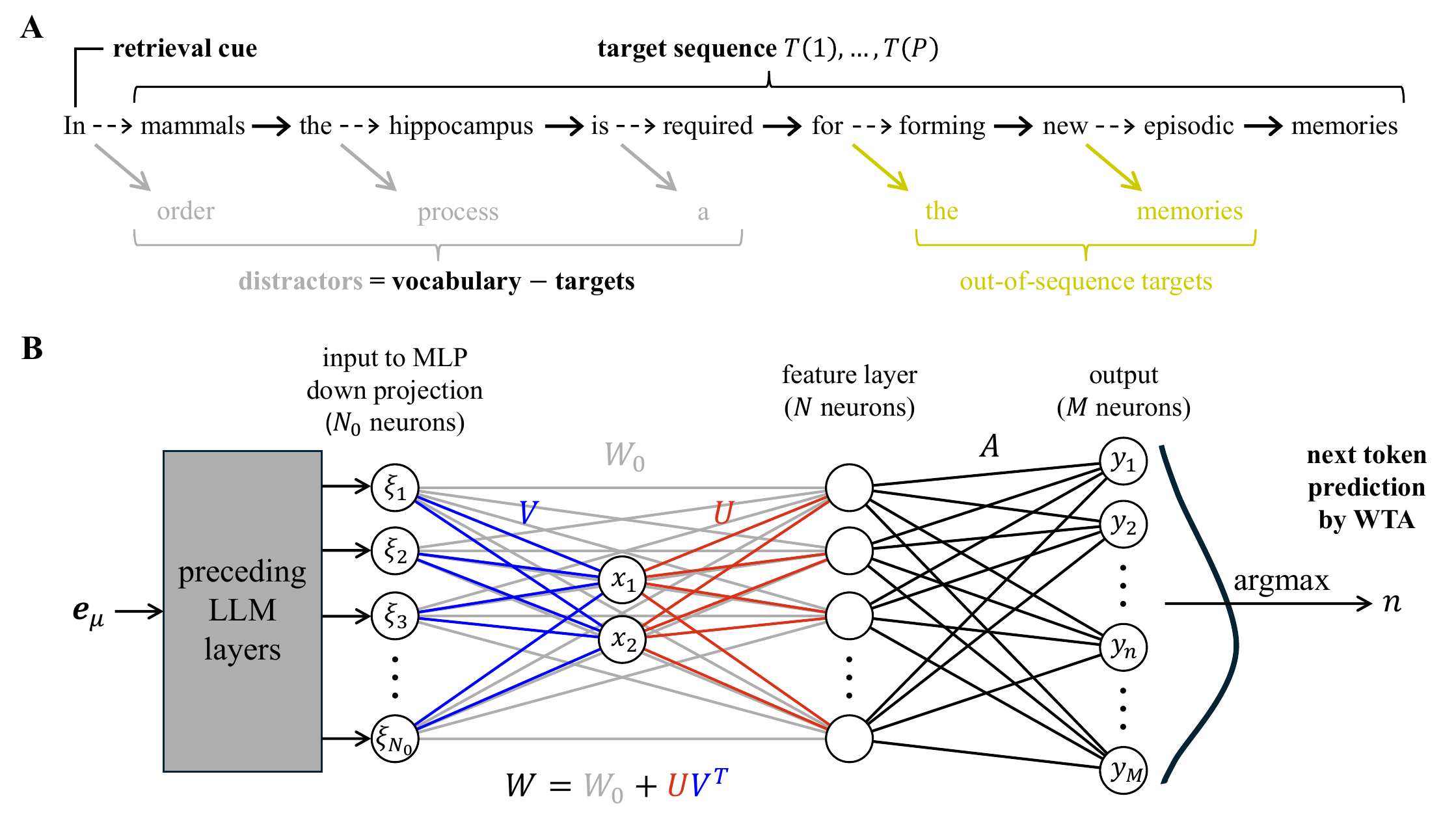}\caption{\textbf{A.} The memorization task. The network is fine-tuned to reproduce a natural target sequence such as the one shown. Before fine-tuning, it makes a number of incorrect associations to tokens that are plausible in context but wrong. If such tokens do not appear anywhere in the target sequence, then they are called \emph{distractors} (gray). Other incorrect predictions are targets that occur elsewhere in the sequence but are inappropriate at the current position (out-of-sequence targets, yellow). The objective of LoRA is to recover the full sequence by suppressing competing predictions and forming the missing associations (dotted arrows). \textbf{B.} Architecture of our LLM-based model. Trainable parameters are shown in color, while fixed parameters and components inherited from the pretrained model are in black or gray.}\label{fig:2}
\end{figure*}

The benefit of pretrained knowledge for new learning has been studied theoretically as transfer learning, focusing on generalization for a target task~\citep{ohzeki2015,ingrosso2025,tahir2024,lauditi2025}, including for LoRA~\citep{duranthon2026}, while other LoRA theories have characterized its expressive power and its training loss landscape~\citep{zeng2024,jang2024}. The storage capacity of full-rank linear associative memories with winner-take-all retrieval has also been analyzed~\citep{giorlandino2026factual,barnfield2026sharp}, but not that of a low-rank modification on an existing network.

Here, we seek to answer three related questions. First, what determines the storage capacity of a low-rank adapter, measured by the maximum episode length~$P$ that it can store? How does this capacity depend on the number of feature neurons~$N$, the latent dimension~$R$, the vocabulary size~$M$, and the episode's semantic content relative to the pretrained network's prior knowledge? Second, what determines the robustness of recall in the presence of noise during retrieval? Third, by what mechanisms does the pretrained base model enhance both storage capacity and recall robustness?

In this paper, we develop a theory that explains the scaling in Fig.~\ref{fig:1}C using geometric arguments and statistical mechanics, and that isolates the important role of the pretrained network's prior knowledge in storing and retrieving episodic memories. A central feature of the architecture is that sequential memory storage becomes a supervised learning problem on many input-output pairs, each mapping the preceding context to the next event. This reformulation avoids many of the instability and interference effects that limit classical recurrent associative memory models. The LoRA connections extract a low-dimensional representation of the input context while suppressing correlations between inputs, and the finite pool of possible events allows winner-take-all competition to select the next one in the sequence. Together, these features permit an analytical treatment of memory storage and retrieval.

We derive the storage capacity and show that it decreases logarithmically with the number of possible events, allowing the vocabulary to vastly exceed the number of neurons at little cost. This weak dependence is essential for realistic episodic memories, in which the number of possible events greatly exceeds the number appearing in any one episode. Memory storage is also highly efficient: for an episode of length~${P}$, each stored association requires~${O(1)}$ trainable parameters, in contrast with the~${O(P)}$ of classical recurrent models. Without the pretrained network, the learned representations and functional connectivity organize into a low-dimensional geometry resembling ring attractor models~\citep{benyishai1995}, while the retrieval margin vanishes as episodes lengthen. The pretrained network changes this picture qualitatively, enabling episodes to be stored with a finite retrieval margin. A solvable model of the pretrained LLM's output reproduces these effects and reveals an emergent form of prediction-error coding in which each association's representation strength grows with its surprise under the pretrained network. This finding is consistent with the observation that hippocampal mismatch signals grow with the strength of the neocortical prediction they violate~\citep{long2016,chen2015}.

\section{The model}\label{sec:model}

We begin with a single episode: a retrieval cue~${T(0)}$ followed by a sequence of~$P$ events~${T(1),\dots,T(P)}$ drawn from an event pool of size~$M$ (Fig.~\ref{fig:2}A). Storing the episode means learning to predict each event from those preceding it. At position~$\mu=1,\dots,P$, the network therefore receives an input~$\boldsymbol{e}_\mu$ consisting of the retrieval cue followed by the first~$\mu-1$ events, and must produce the next event~$T\left(\mu\right)$. We call these desired outputs \emph{targets}, which need not be distinct, and events that never appear among~$T(1),\dots,T(P)$ \emph{distractors}. At position~$\mu$, every event~$n\neq T(\mu)$ is a \emph{competitor}, including targets at other positions (\emph{out-of-sequence} targets). Thus, competitor is a position-dependent role, whereas membership in the target or distractor set is defined for the whole episode.

We model memory formation as LoRA fine-tuning of a pretrained LLM, in which events are tokens and the event pool is the model’s vocabulary. For analytical tractability, we restrict fine-tuning to the down-projection of the final multilayer perceptron (MLP) block (Fig.~\ref{fig:2}B). The preceding layers map~$\boldsymbol{e}_\mu$ to a representation ${\boldsymbol{\xi}\left(\boldsymbol{e}_\mu\right)\in\mathbb{R}^{N_0}}$. Then the down-projection maps this representation to an $N$-dimensional feature layer, and a readout~$A\in\mathbb{R}^{M\times N}$ projects the features onto~$M$ output units, among which winner-take-all (WTA) selects the output token. During recall, the selected token is fed back as part of the context, generating the next prediction and thereby reconstructing the sequence autoregressively.

The pretrained network, which includes the preceding layers, the pretrained down-projection~$W_0$, and the readout matrix, is kept frozen. The down-projection after fine-tuning is
\begin{equation}
W=W_{0}+UV^{T},
\end{equation}
where~$U\in\mathbb{R}^{N\times R}$ and~$V\in\mathbb{R}^{N_{0}\times R}$ define a rank-$R$ LoRA update, with~$R\ll N,N_{0}$. Writing~$\boldsymbol{a}_{n}$ for the readout vector of output unit~$n$, the output takes the form
\begin{equation}
y_{n}\left(\boldsymbol{e}_\mu\right)=b_{n}\left(\boldsymbol{e}_\mu\right)+\boldsymbol{h}_{n}\cdot\boldsymbol{x}\left(\boldsymbol{e}_\mu\right),
\end{equation}
where~$\boldsymbol{b}\left(\boldsymbol{e}_\mu\right)=AW_{0}\boldsymbol{\xi}\left(\boldsymbol{e}_\mu\right)$ is the contribution of the pretrained network,~$\boldsymbol{x}\left(\boldsymbol{e}_\mu\right)=V^{T}\boldsymbol{\xi}\left(\boldsymbol{e}_\mu\right)$ is the learned $R$-dimensional latent representation of the input, and~${\boldsymbol{h}_{n}=U^{T}\boldsymbol{a}_{n}\in\mathbb{R}^R}$ is the functional connection from the latent to output unit~${n}$, scaled without loss of generality so that~$\max_{n}\left|\boldsymbol{h}_{n}\right|=1$.

Storing the sequence for deterministic recall requires each target token to win the WTA competition, $\text{argmax}_{n}\left\{ y_{n}\left(\boldsymbol{e}_\mu\right)\right\} =T\left(\mu\right)$. To quantify the robustness of recall, we require each one to win by a margin~${\kappa|\boldsymbol{x}_\mu|}$, where~$\kappa>0$ and the factor of~${|\boldsymbol{x}_\mu|}$ prevents the norms of the learned latent representations from diverging when~${\kappa}$ is maximized. Dividing by~${|\boldsymbol{x}_\mu|}$, the WTA constraints become
\begin{equation}
\label{eq:3}
\boldsymbol{\hat{x}}_\mu\cdot\left(\boldsymbol{h}_{T\left(\mu\right)}-\boldsymbol{h}_{n}\right)+\left|\boldsymbol{x}_\mu\right|^{-1}\delta^n_\mu\geq\kappa>0
\end{equation}
for all~${\mu=1,\dots,P}$ and~${n\neq T(\mu)}$. Here,~$\boldsymbol{\hat{x}}_\mu=\boldsymbol{x}_\mu/\left|\boldsymbol{x}_\mu\right|$ are the latent directions, and~$\delta^{n}_\mu=b^{T\left(\mu\right)}_\mu-b^{n}_\mu$ is the pretrained model's \emph{relative bias} toward the correct target token~$T\left(\mu\right)$ over competitor~$n\neq T(\mu)$. Throughout, relative biases and extrema over them refer only to competitors. Positive values favor the target; negative values favor the competitor. The first term is the learned low-rank contribution, which admits a simple geometric picture in~${R}$ dimensions that we will exploit throughout the paper. The second is the pretrained model's output, whose weight relative to the learned term is set by~${\left|\boldsymbol{x}_\mu\right|}$.

To train the network, we take~$P\leq N_0$ and the input representations~$\boldsymbol{\xi}_\mu=\boldsymbol{\xi}\left(\boldsymbol{e}_\mu\right)$ to be linearly independent, so that~${V}$ can realize any set of latent representations~$\boldsymbol{x}_\mu$. We therefore first optimize~${\left\{\boldsymbol{x}_\mu\right\}}$ together with~$U$ and then recover~$V$ by linear regression. Since the second stage is linear, the essential memory storage problem lies entirely in the first stage. For the analysis, we take the entries of~$A$ to be i.i.d. Gaussian with zero mean and variance~${1/N}$. Despite this simplification, the resulting capacity predictions agree with simulations using real LLM readouts in the most general version of our model (Appendix D).

For the statistical mechanical analysis of this model, we take the following thermodynamic limit: $N,P,M\to\infty$, with the load~$\alpha=P/N$, the vocabulary load~$\beta=M/N$, and the rank~$R$ held fixed. A challenge in applying this limit is that Eq.~\eqref{eq:3} imposes~$P(M-1)=O(N^2)$ inequalities. As we show below, a combination of symmetry considerations and simplifying assumptions about the structure of the pretrained relative biases allows us to reduce the problem to a number of constraints linear in~$N$ and characterize the storage capacity, retrieval margin, and solution geometry of the LoRA-only and pair bias models.

\begin{figure}
\centering{}\includegraphics[width=\columnwidth]{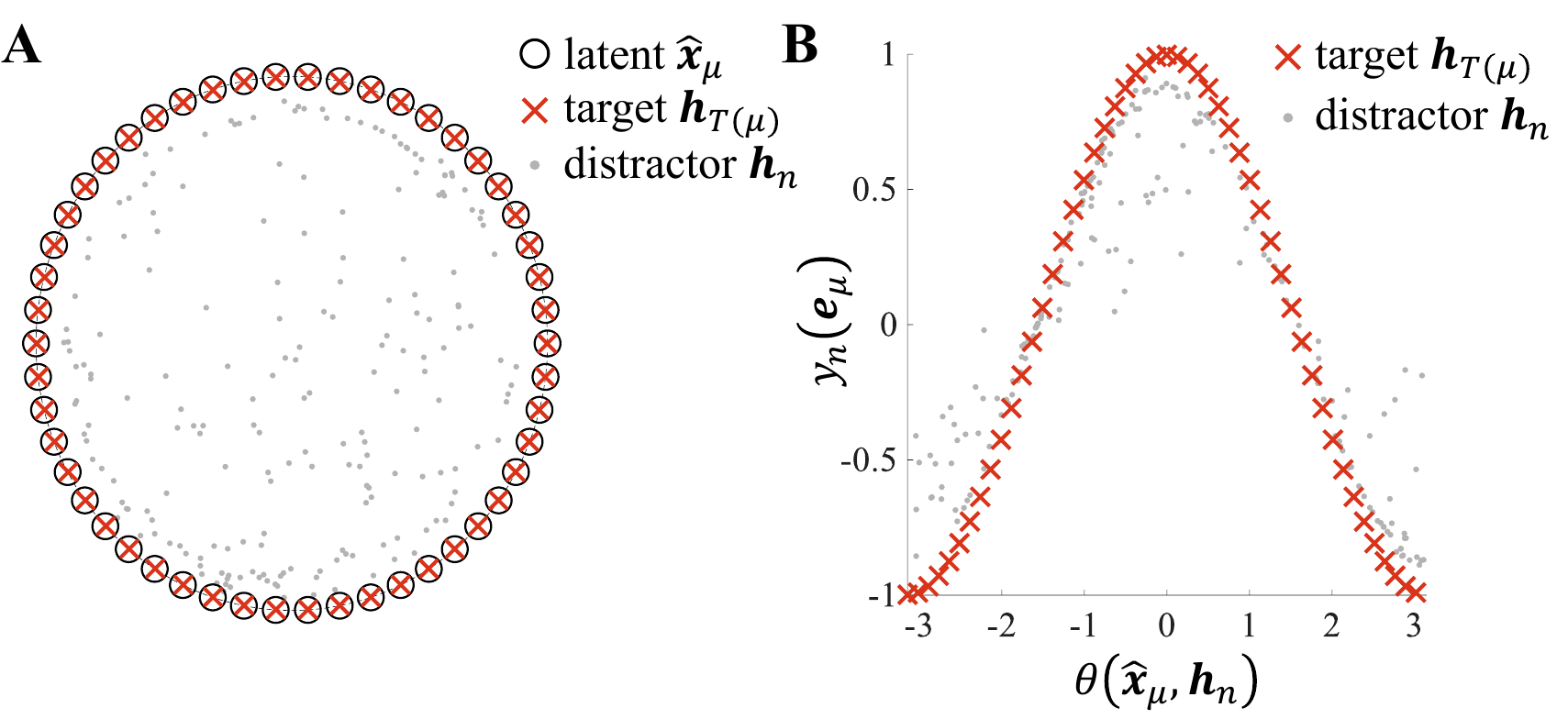}\caption{\textbf{A.} The maximum-margin solution of the LoRA-only model for~${R=2}$ (other parameters:~$N=130$, $P=50$, $M=250$). The learned vectors~$\boldsymbol{h}_{T\left(\mu\right)}$ align with the corresponding latent directions~$\boldsymbol{\hat{x}}_{\mu}$
on the unit circle, while the distractor vectors~$\boldsymbol{h}_{n}$ remain broadly distributed in the unit disk, in agreement with Eq.~\eqref{eq:10}. \textbf{B.} The logit distribution produced by the input~$\boldsymbol{e}_{\mu}$ for the solution in A.
}\label{fig:3}
\end{figure}

\section{LoRA-only}\label{sec:LoRA-only}

\begin{figure*}[t]
  \centering\includegraphics[width=1\textwidth]{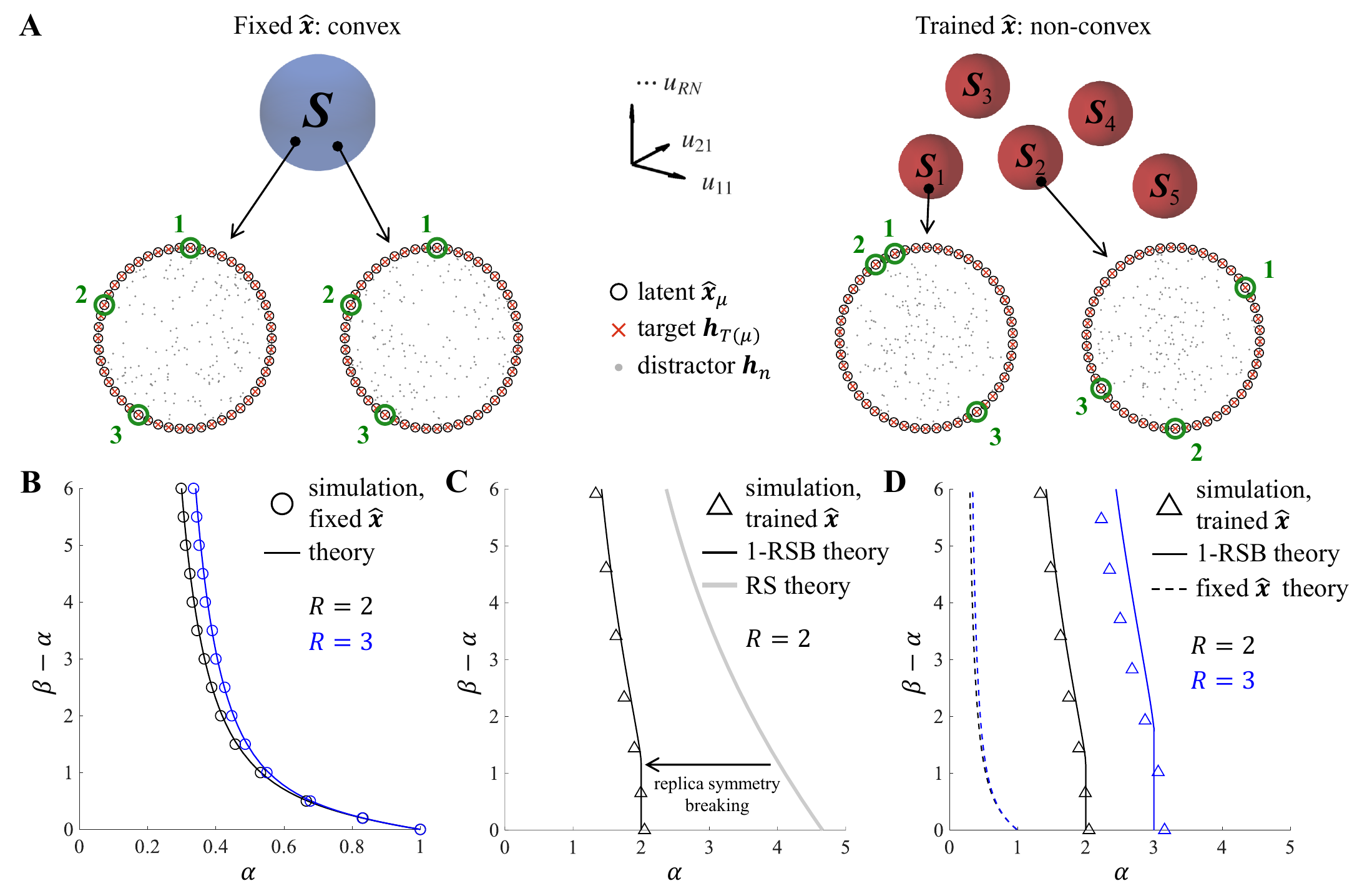}\caption{\textbf{A.} \textbf{Top:} Illustration of solution regions in weight ($U$) space near capacity. For fixed~$\boldsymbol{\hat{x}}$, the region is convex and hence connected, and it shrinks as capacity is approached, so any two solutions lie close together. For trained~$\boldsymbol{\hat{x}}$, the region fragments into disconnected pieces, so two solutions at capacity can lie far apart. \textbf{Bottom:} Two randomly sampled pairs of maximum-margin solutions with identical readouts~${A}$ (same parameters as in Fig.~\ref{fig:3}A). Green circles mark the functional connections~$\boldsymbol{h}_{1},\boldsymbol{h}_{2},\boldsymbol{h}_{3}$. With fixed~$\boldsymbol{\hat{x}}$ (RS), the target connections have identical configurations in the two solutions. With trained~$\boldsymbol{\hat{x}}$ (RSB), their relative positions and clockwise ordering differ. In both cases, the distractor connections do not set the margin and are free to move inside the disk. \textbf{B.} Normalized distractor set size~$\beta-\alpha$ versus capacity~$\alpha$ for fixed~$\boldsymbol{\hat{x}}_{\mu}$ in simulations versus theory. \textbf{C.} Normalized distractor set size~$\beta-\alpha$ versus capacity~$\alpha$ for trained~${\boldsymbol{\hat{x}}_{\mu}}$ in simulations versus RS and 1-RSB theories ($R=2$). \textbf{D.} The trained~$\boldsymbol{\hat{x}}$ capacity (solid lines, triangles) increases roughly proportional to~$R$, and is upper-bounded by~$\alpha=R$, a substantial increase over the fixed~$\boldsymbol{\hat{x}}$ capacity (dashed lines). Panels B-D use~${P=600}$ for~${R=2}$, and~${P=6000}$ for~${R=3}$ (due to slower convergence to the thermodynamic limit at larger~${R}$).
}\label{fig:4}
\end{figure*}

We first consider the LoRA-only case where~$W_0=0$ and the relative bias vanishes, $\boldsymbol{\delta}=0$. The storage constraints~\eqref{eq:3} reduce to
\begin{equation}
\label{eq:5}
\boldsymbol{\hat{x}}_\mu\cdot\left(\boldsymbol{h}_{T\left(\mu\right)}-\boldsymbol{h}_n\right)\geq\kappa>0
\end{equation}
for every position~$1 \leq \mu \leq P$ and every competitor ${n\neq T\left(\mu\right)}$.

In the thermodynamic limit, these constraints determine the target functional connections, as we show with a geometric argument (Appendix A):
\begin{equation}
\label{eq:10}
\boldsymbol{h}_{T\left(\mu\right)}\rightarrow\boldsymbol{\hat{x}}_\mu,\quad\kappa\rightarrow0 ,
\end{equation}
while the remaining distractor connections~$\boldsymbol{h}_n$ are confined to the interior of the unit ball,~${\left|\boldsymbol{h}_n\right|<1}$. Simulations with~${R=2}$ confirm this picture once~$P$ and~$M$ are large (Fig.~\ref{fig:3}A). As a result, the target logits~${y_n(\boldsymbol{e}_\mu)}$ for any given~$\mu$ follow a cosine distribution in the angle~${\theta\left(\boldsymbol{\hat{x}}_\mu,\boldsymbol{h}_n\right)}$, with the peak occurring at~${n=T\left(\mu\right)}$ for which~$\boldsymbol{h}_n$ and~$\boldsymbol{\hat{x}}_\mu$ are maximally aligned, while the distractor logits are strictly less in absolute value (Fig.~\ref{fig:3}B).

Equation~\eqref{eq:10} leaves open how the latent directions~${\boldsymbol{\hat{x}}_\mu}$ are chosen. In LoRA fine-tuning, and in all of our LLM experiments, they are learned jointly with~${U}$. For the theoretical analysis, we first consider the simpler case in which they are fixed in advance, before turning to this full problem (Fig.~\ref{fig:4}A). Comparing the two cases also shows how much is gained by learning the latent directions.

\paragraph{Fixed~$\boldsymbol{\hat{x}}_\mu$.}
With the latent directions fixed, the constraints~$\boldsymbol{h}_{T\left(\mu\right)}=\boldsymbol{\hat{x}}_\mu$ are convex in the weights~$U$. We use Gardner's method~\citep{gardner1988} and the replica-symmetric (RS) ansatz to average over the quenched weights~$A$ and compute the storage capacity~$\alpha=P/N$. Writing~$\beta=M/N$ for the vocabulary load, the resulting mean-field equations are
\begin{equation}
\label{eq:1}
Rq-\alpha\left(Rq+1\right)-\left(\beta-\alpha\right)\left\langle \left[\sqrt{q}\left|\boldsymbol{z}\right|-1\right]^{2}_{+}\right\rangle _{\boldsymbol{z}}=0,
\end{equation}
\begin{equation}
\label{eq:2}
R\sqrt{q}-R\alpha\sqrt{q}-\left(\beta-\alpha\right)\left\langle \left|\boldsymbol{z}\right|\left[\sqrt{q}\left|\boldsymbol{z}\right|-1\right]_{+}\right\rangle _{\boldsymbol{z}}=0,
\end{equation}
where~$\boldsymbol{z}$ is an $R$-dimensional Gaussian vector,~$\left[\cdot\right]_{+}$ the ReLU function, and~${q=\frac{1}{RN}\left|U\right|^{2}_{F}}$ the squared norm of the LoRA weights. Their numerical solution is in excellent agreement with simulations (numerical methods in Appendix~\ref{app:methods}) at large~$N$, $P$, and~$M$ (Fig.~\ref{fig:4}B).

\paragraph{Trained~$\boldsymbol{\hat{x}}_\mu$.}
Next, we turn to the full problem, in which the latent directions are learned jointly with~$U$. Equation~\eqref{eq:10} now supplies the latent directions themselves, $\boldsymbol{\hat{x}}_\mu=\boldsymbol{h}_{T(\mu)}$, so that the only surviving storage constraints are
\begin{equation}
\label{eq:7}
\left|\boldsymbol{h}_{T\left(\mu\right)}\right|=1,
\end{equation}
in addition to the distractor constraints,~$\left|\boldsymbol{h}_{n}\right|<1$. Equation~\eqref{eq:7} is nonconvex, resulting in fragmentation of the solution space as in the committee machine~\citep{barkai1992}, which we illustrate in Fig.~\ref{fig:4}A. Numerical optimization is consequently not guaranteed to find the global optimum, and replica symmetry need no longer hold. We demonstrate this by computing both the RS and one-step replica symmetry breaking (1-RSB)~\citep{mezard1987} solutions (capacity equations are given in Appendix C). Figure~\ref{fig:4}C shows that the 1-RSB solution matches simulations substantially better than the RS, indicating genuine replica symmetry breaking. The residual gap between theory and simulation for~${R=3}$ reflects either further replica symmetry breaking or the sub-optimality of the numerical optimization for the nonconvex problem, which increases in difficulty with larger~${R}$ (Fig.~\ref{fig:4}D). The maximum capacity increases from~$1$, in the case of fixed~$\boldsymbol{\hat{x}}_\mu$, to~$R$. We now provide a more intuitive argument for these two different upper bounds.

For fixed~$\boldsymbol{\hat{x}}_\mu$, the capacity at~$\beta=\alpha$ admits an elementary derivation requiring no replica calculation. The constraints~$\boldsymbol{h}_{T\left(\mu\right)}=\boldsymbol{\hat{x}}_\mu$ form a system of~$PR$ scalar equations in~$NR$ unknowns, so at capacity
\begin{equation}
PR=NR\Rightarrow\alpha=1.
\end{equation}
For trained~$\boldsymbol{\hat{x}}_\mu$, the weaker constraints~$\left|\boldsymbol{h}_{T\left(\mu\right)}\right|=1$ give only~$P$ equations in~$NR$ unknowns, so
\begin{equation}
P=NR\Rightarrow\alpha=R,
\end{equation}
provided the intersection of the~$P$ cylindrical regions defined by the constraints is nonempty. Each bound is the maximum capacity attainable at the smallest vocabulary,~$\beta=\alpha$.

The large-$\beta$ regime is especially relevant to modern LLMs. For instance, Llama-3.1-8B-Instruct, which we use in our LLM simulations, has a vocabulary of~${M=\text{128,256}}$ tokens, orders of magnitude larger than the length of the sequences considered here. Even a modest LoRA with~${N=800}$ feature neurons is enough to memorize them and gives~${\beta\approx160}$. In this regime, the mean-field equations yield the same capacity to leading order for fixed~$\boldsymbol{\hat{x}}$ and trained~$\boldsymbol{\hat{x}}$ (including both the RS and 1-RSB predictions):
\begin{equation}
\label{eq:6}
\alpha\approx\frac{R}{2\log\beta},
\end{equation}
up to corrections that decay slowly with~${\log\beta}$. Thus, the capacity decreases only logarithmically with vocabulary size, independently of whether the latent directions are learned. This weak dependence is what makes the model viable for realistic sequential memories, where the vocabulary can vastly exceed the sequence length. We show later that the same logarithmic dependence on vocabulary and linear dependence on~${R}$ persist when we reintroduce the base connectivity, suggesting they may be general features of WTA retrieval through a low-rank bottleneck.

\begin{figure}
\centering{}\includegraphics[width=\columnwidth]{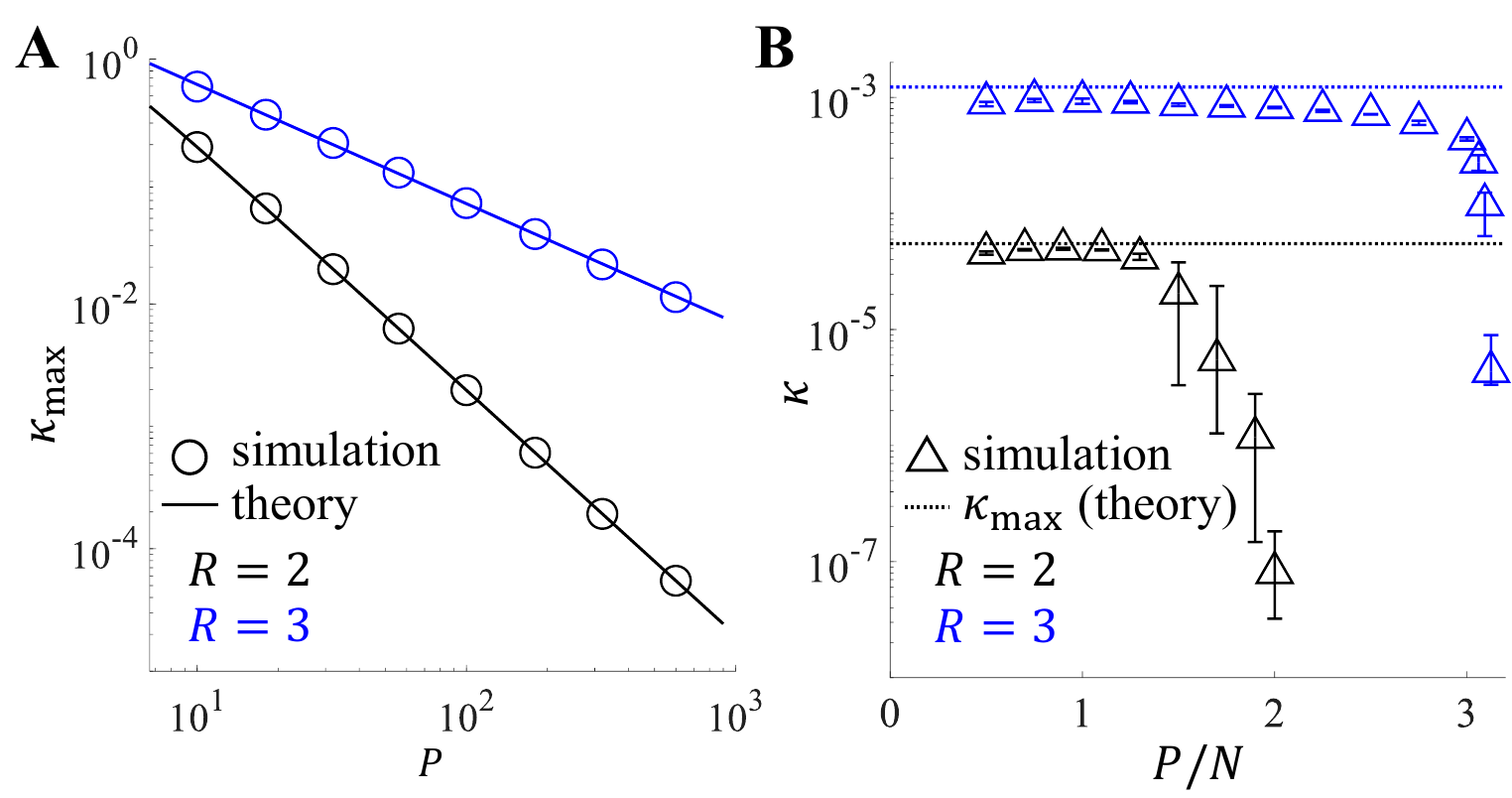}\caption{\textbf{A.} Maximal margin versus sequence length~${P}$ for fixed~$\boldsymbol{\hat{x}}$ (circles, simulations; lines, the theoretical scaling of Eq.~\eqref{eq:4}). \textbf{B.} Margin versus load~${P/N}$ for trained~$\boldsymbol{\hat{x}}$, with $P=M=600$ and decreasing~$N$. The margin decreases smoothly from~${\kappa_{\max}}$ (dashed lines) to~$0$ as~${P/N}$ approaches the capacity~${\alpha=R}$. Triangles show the geometric mean of~${\kappa}$ over~4 runs, and error bars the geometric standard deviations above and below it, computed separately.}\label{fig:5}
\end{figure}

\begin{figure*}[t]
\centering\includegraphics[width=0.9\textwidth]{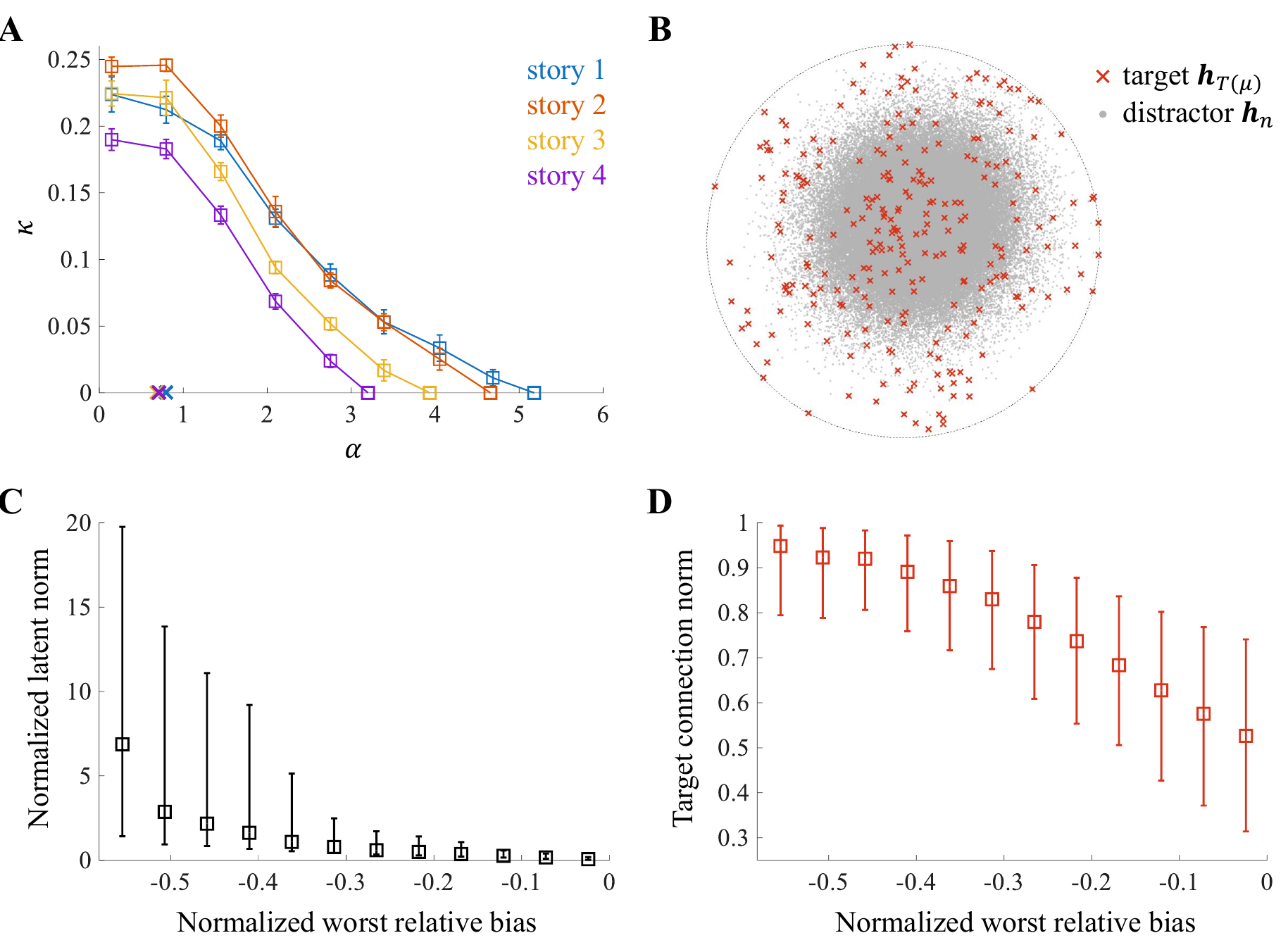}
\caption{Full-model solutions. \textbf{A.} Margin~$\kappa$ versus capacity~$\alpha$ in the fine-tuned LLM for four stories of length~${P=600}$. Capacity is modulated by varying~$N$, and error bars show the standard deviation across~10 trials using different random subsets of the feature layer and weight initializations. Crosses indicate the theoretical LoRA-only capacity and maximal margin for the same stories. \textbf{B.} Functional connections in the fine-tuned LLM. \textbf{C.} Minimal latent representation norms along the latent directions found by the optimizer, versus the worst relative biases~${\min_n\delta^n_\mu}$ for all targets~$\mu$ not predicted by the base model, with both quantities normalized by the mean relative bias~$\bar\delta$. Here~$\bar\delta$ is the average of~$\delta^n_\mu$ over all positions~$\mu$ and competitors. \textbf{D.} Target functional connection norms versus the same normalized worst relative biases. In C and D, targets are pooled over~104 stories, and error bars show the standard deviations of the values above and below the mean. All panels use~${R=2}$ and the full LLM vocabulary~${M=128{,}256}$. Panels B--D use the LLM's maximum~${N=4096}$.}\label{fig:6}
\end{figure*}

\paragraph{The margin.}
Robust memory retrieval requires that the inputs to the target units are separated from all others by a substantial margin. On the other hand, in the thermodynamic limit studied above the margin vanishes. Here, we ask how quickly the maximal margin approaches~0 as~$P\rightarrow \infty$. For~$R=2$, the maximum is achieved when the latent directions form a regular lattice on the unit circle, giving~$\kappa_{\max}\approx2\pi^{2}/P^{2}$. More generally,
a simple packing argument (see Appendix A) gives the following scaling illustrated in Fig. \ref{fig:5}A:
\begin{equation}
\label{eq:4}
\kappa_{max}\sim P^{-\frac{2}{R-1}},
\end{equation}
where the precise coefficient is related to the solution of the sphere-packing problem~\citep{conway1999} in $R-1$ dimensions. For fixed~$\boldsymbol{\hat{x}}_{\mu}$, where~$\alpha\leq1$, this optimum is always attainable because the latent directions can be chosen to be evenly distributed on the sphere. But for trained~$\boldsymbol{\hat{x}}_{\mu}$, this is not the case. Simulations at~$\beta=\alpha$ show that once~${\alpha}$ exceeds~1, the margin decreases continuously to zero as~${\alpha\rightarrow R}$, even for finite~${P}$ (Fig.~\ref{fig:5}B).

Repeated target tokens do not require any separate treatment, since two positions with the same target impose the same constraint on~${U}$. The asymptotic solutions~\eqref{eq:10} and~\eqref{eq:7} hold unchanged, with the single stipulation that for fixed~${\boldsymbol{\hat{x}}}_\mu$ two positions sharing target must be assigned the same latent direction, while for trained~${\boldsymbol{\hat{x}}_\mu}$ this property arises of its own accord. Duplicate associations therefore consume no additional feature neurons, and the capacity conditions derived above hold with~${P}$ replaced by the number of distinct targets,~${P_{\text{unique}}}$.

Adding the pretrained connectivity~$W_0$ changes this. Two positions with the same target can impose different constraints on~${U}$ due to their different biases, so their optimal latent directions need not be the same. As a result, the relevant capacity measure will be~${\alpha=P/N}$. As we show in the next section, a more striking consequence of the pretrained model is an~${O(1)}$ margin, ${\kappa_{\max}\sim0.2}$ at~${P=600}$, and the pair model of Sec.~V explains why a margin of this order persists in the thermodynamic limit.

\begin{figure*}[t]
\centering\includegraphics[width=\textwidth]{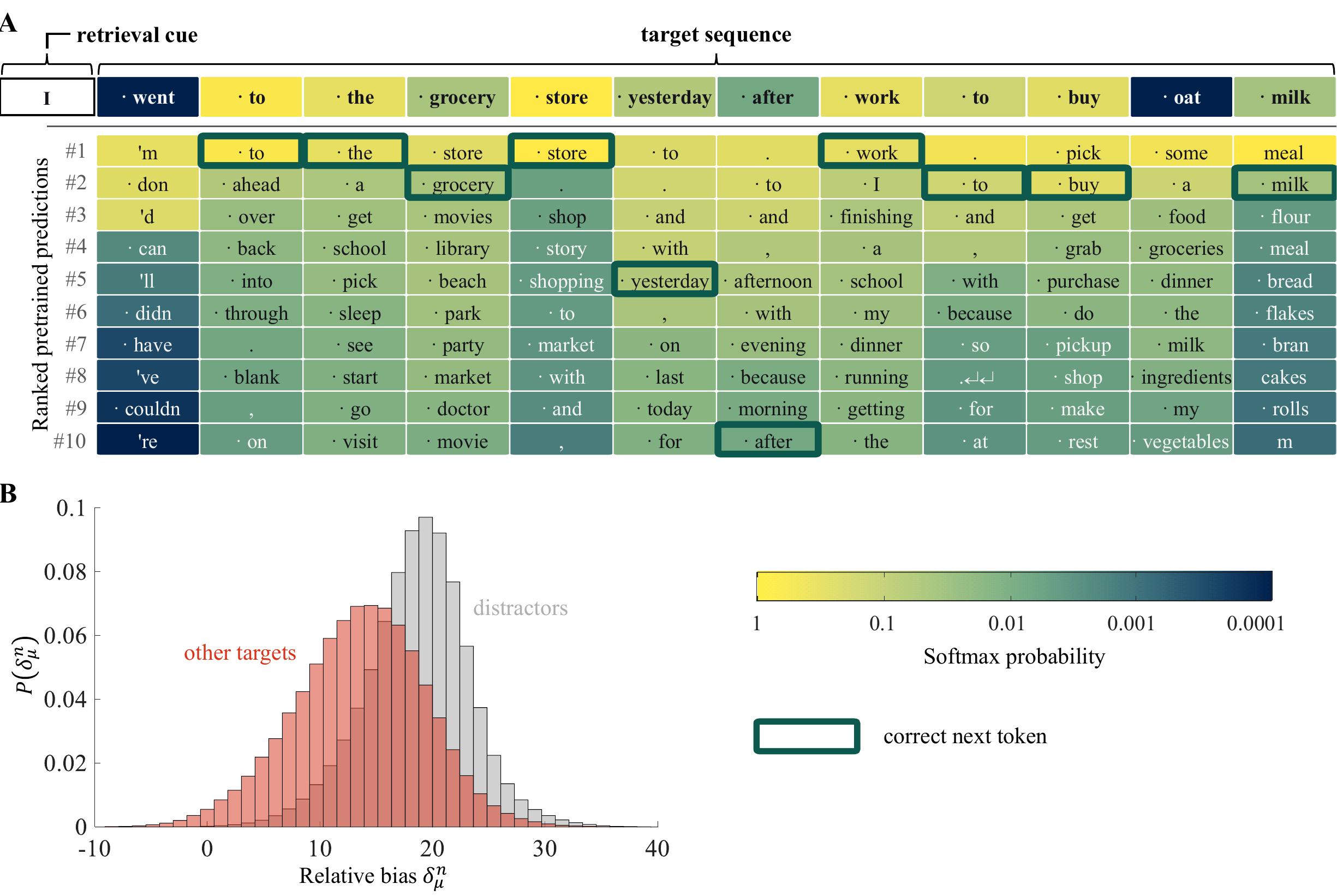}
\caption{Relative biases underlying the full-model solutions. \textbf{A.} Ranked predictions of the pretrained LLM on an example sequence. Dots (``$\cdot$'') indicate spaces, and~\Pisymbol{psy}{191} a line break. \textbf{B.} Distribution of the relative biases~$\delta_\mu^n$ defined in Eq.~\eqref{eq:3}, for competitors that appear as targets elsewhere in the sequence (red) and for distractors, which do not appear in the sequence at all (gray).}\label{fig:7}
\end{figure*}

\section{Fine-tuning with LLM base connectivity}\label{sec:full_model}

We now restore the pretrained connectivity~${W_0}$ and examine how it changes the solutions found by LoRA fine-tuning. We use stories of length~${P=600}$ and the full LLM vocabulary,~${M=128{,}256}$. To vary the load~${\alpha=P/N}$, we restrict the adapter to random subsets of~${N}$ of the LLM's~4096 feature neurons, while the pretrained relative biases~${\delta_\mu^n}$ in Eq.~\eqref{eq:3} are computed with the full network. Different random subsets give the trial-to-trial variability.

The pretrained connectivity substantially increases both storage capacity and retrieval margin (Fig.~\ref{fig:6}A). The capacity rises from~${\alpha\approx0.8}$ in the LoRA-only model to~${\alpha\approx3}$--$5$ in the full model. The maximal margin increases from~${\sim2\times10^{-4}}$, the LoRA-only regular-lattice value~${2\pi^2/P_{\mathrm{unique}}^2}$ for~${P_{\mathrm{unique}}\sim300}$ distinct targets, to~${\kappa_{\max}=0.22\pm0.02}$. Thus, at the sequence lengths studied here, the full model combines a much larger capacity with a substantial retrieval margin.

The solutions also differ geometrically from those of the LoRA-only model. Using all~4096 feature neurons, we find that the target functional connections~${\boldsymbol{h}_{T(\mu)}}$ occupy the interior of the unit disk rather than concentrating on its boundary (Fig.~\ref{fig:6}B). The latent representation norms~${|\boldsymbol{x}_\mu|}$ vary broadly across positions and grow as the worst relative bias~${\min_n\delta_\mu^n}$ becomes more negative (Fig.~\ref{fig:6}C). For these targets, which the pretrained network predicts incorrectly, the negative worst relative bias is the zero-temperature limit of the temperature-scaled surprisal, $-\min_n\delta^n_\mu=\lim_{T\to0}T\,[-\log p_T]$, where $p_T$ is the probability the pretrained network assigns to the correct target at softmax temperature $T$. The norms therefore grow with the pretrained network’s surprise in this limit. The target functional connection norms follow the same trend (Fig.~\ref{fig:6}D): the most unfavorable biases are associated with vectors closer to the unit circle, approaching the LoRA-only geometry. The dependence of both amplitudes on the pretrained relative biases is therefore a further salient feature of the full-model solutions.

What properties of the pretrained relative biases account for these changes? Consider the example sequence in Fig.~\ref{fig:7}A. The pretrained LLM already predicts several of its tokens correctly, and even when its top prediction is wrong, the desired token is often ranked highly. Most competitors are consequently disfavored relative to the correct target: the overwhelming majority of~${\delta_\mu^n}$ are positive, with negative biases confined to a sparse subset of competitors (Fig.~\ref{fig:7}B).

\begin{figure*}[t]
\centering\includegraphics[width=0.9\textwidth]{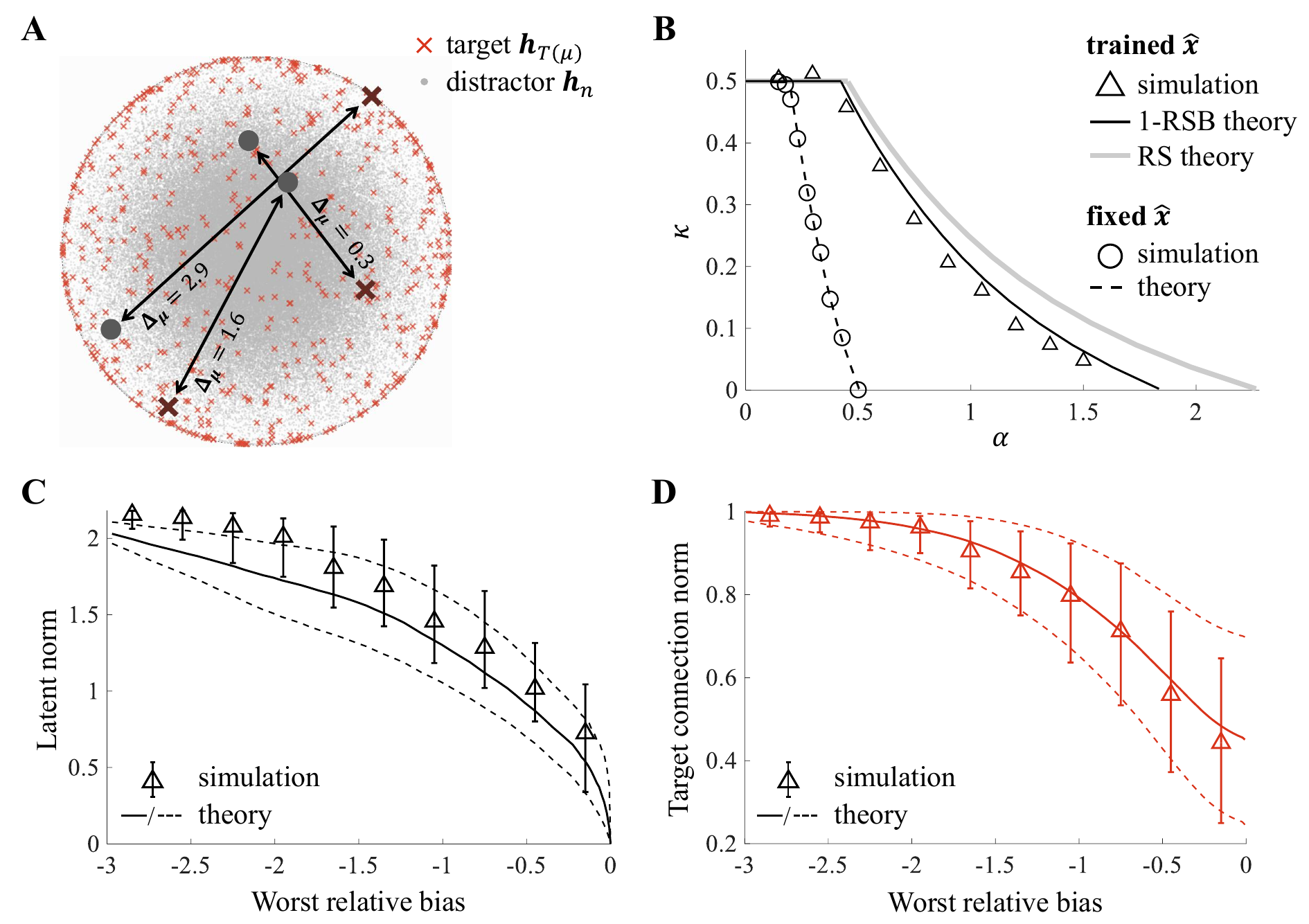}
\caption{Pair-model solutions. \textbf{A.} Functional connections in the pair model. Gray circles denote distractors, including rivals. Arrows connect a few example targets (enlarged red crosses) to their rivals (enlarged gray circles), showing that pairs with larger~${\Delta_\mu}$ are pushed further apart. \textbf{B.} Margin versus capacity, with capacity modulated by varying~$N$, for trained and fixed latent directions (simulations, symbols; theory, lines). \textbf{C.} Latent representation norms versus the worst relative biases~${\min_n\delta_\mu^n=-\Delta_\mu}$, as in Fig.~\ref{fig:6}C but without normalization since~${\bar{\delta}\approx1}$, and with latent directions drawn uniformly from the feasible set. Symbols and lines show simulations and theory, respectively, with error bars and dashed lines giving the standard deviations above and below the mean. \textbf{D.} Target functional connection norms versus the worst relative biases, shown as in C. All panels use~${P=600}$ unique targets, the full LLM vocabulary~${M=128{,}256}$, and~$p(\Delta)$ uniform on~${[0,3]}$. Panels A, C, and D use~${N=1342}$, for which~${\kappa}$ sits just below its maximum.}\label{fig:8}
\end{figure*}

This structure provides a qualitative explanation of the increased capacity and margin. In Eq.~\eqref{eq:3}, positive biases assist the adapter in satisfying the corresponding storage constraints; only the unfavorable competitors require the adapter to overcome the pretrained output. The learned functional connections therefore need not maintain the tightly packed spherical configuration of the LoRA-only solution. The same constraints explain why the learned amplitudes vary with the biases: overcoming a more negative bias requires a larger latent norm to reduce its relative contribution, together with a sufficiently large projected separation between the target and its competitors. These observations motivate a simpler model that retains the predominance of favorable biases and the sparse unfavorable competitions.

\section{Pair bias model}\label{sec:pair}

In the pair bias model, the relative bias for each target's single \emph{rival}~$R(\mu)$, a designated competitor, is~$\delta_\mu^{R(\mu)}=-\Delta_\mu<0$. For all other competitors,~$\delta_\mu^n=+1$. Thus, the base model favors the rival over the target and the target over all other competitors. This sparse pattern of unfavorable relative biases defines a tractable model intended to reproduce the main features of the full-model solutions: a substantial margin and increased capacity, functional connections inside the unit sphere, and latent and connection amplitudes that increase with unfavorable relative biases.

For simplicity, we take the target tokens to be distinct and choose a different rival from the distractor pool for each target. Rivals are therefore a subset of distractors in this model; all other competitors include both the remaining distractors and the other targets. We draw~$\Delta_{\mu}>0$ independently from a distribution~$p\left(\Delta\right)$, giving
\begin{equation}
\delta^{n}_{\mu}=\begin{cases}
-\Delta_{\mu}, & n=R\left(\mu\right)\\
1, & \text{otherwise}
\end{cases}
\end{equation}
The negatively biased target-rival pairs are therefore disjoint. The analysis below extends straightforwardly to~$k=O\left(1\right)$ rivals per target, provided they are not shared between targets, but the pair model already captures the main qualitative features observed in the LLM.

Optimizing over~$\left|\boldsymbol{x}_{\mu}\right|$, the~${M-1}$ WTA constraints for each target collapse in the thermodynamic limit into a single constraint (see Appendix A):
\begin{equation}
\boldsymbol{\hat{x}}_{\mu} \cdot \left( \left( 1 + \Delta_\mu \right) \boldsymbol{h}_{T\left(\mu\right)} - \boldsymbol{h}_{R\left(\mu\right)} \right) \geq \Delta_{\mu}\kappa + \Delta_{\mu} + \kappa > 0.\label{eq:15}
\end{equation}

Figure~\ref{fig:8}A illustrates how the model resolves these competitions. The learned target and rival connections separate along the latent direction, and pairs with larger~${\Delta_\mu}$ require a stronger correction. The relevant separation is its projection along~${\boldsymbol{\hat{x}}_\mu}$, which must overcome the unfavorable bias while leaving a positive retrieval margin. Unlike the LoRA-only constraints, Eq.~\eqref{eq:15} allows target connections to lie inside the unit ball.

The maximal margin of a pair is attained when~${\boldsymbol{h}_{T(\mu)}=\boldsymbol{\hat{x}}_\mu}$ and~${\boldsymbol{h}_{R(\mu)}=-\boldsymbol{\hat{x}}_\mu}$. For the most unfavorable pair this gives
\begin{equation}
\label{eq:pair_kappa_max}
\kappa_{\max}=\frac{2}{1+\Delta_{\max}},
\end{equation}
where~${\Delta_{\max}}$ is the largest value in the support of~${p(\Delta)}$. For a bounded bias distribution, the model therefore supports an~${O(1)}$ margin in the thermodynamic limit.

The model can be solved analytically using replica mean-field theory, with the resulting capacity equations evaluated numerically.

For fixed latent directions~$\boldsymbol{\hat{x}}_{\mu}$ on the sphere, the optimization is convex and replica symmetric. The resulting mean-field equations are given in Appendix~B, and their numerical solution agrees closely with simulations at large~$N$, $P$, and $M$. When the latent directions are trained jointly with the functional connections, the problem becomes nonconvex and replica symmetry breaking occurs. In this case, 1-RSB provides a better description of the simulations than RS (see Fig.~\ref{fig:8}B and mean-field equations given in Appendix C). The agreement persists when the Gaussian readout in the simulations is replaced by the readout layers of real LLMs (Appendix D).

The margin-capacity curves in Fig.~\ref{fig:8}B reproduce the qualitative behavior of the full LLM in Fig.~\ref{fig:6}A: a finite maximal margin at low load, followed by a decline toward zero as capacity is approached. The capacity values need not coincide, since the synthetic bias distribution and distinct-target construction differ from those of the full LLM.

\begin{figure}
\centering\includegraphics[width=0.8\columnwidth]{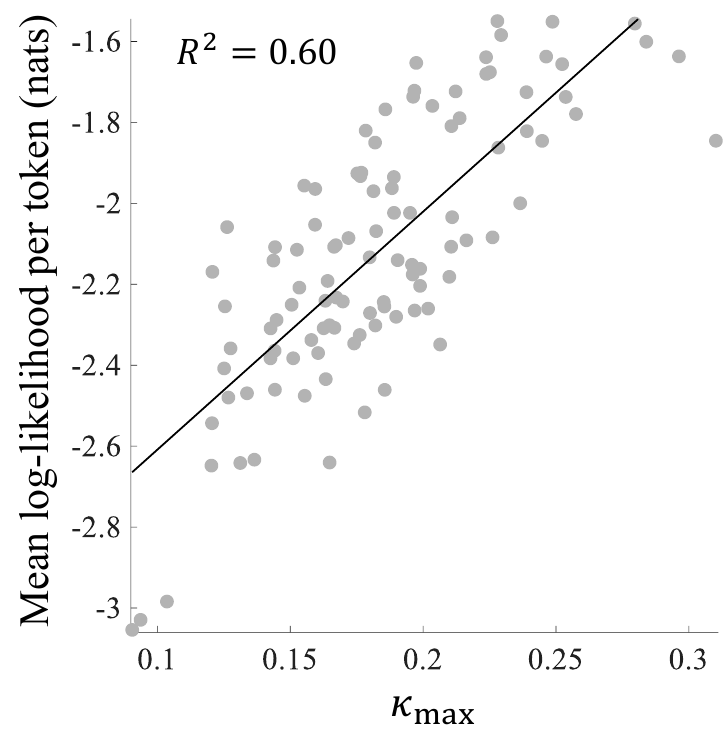}
\caption{Story predictability and the maximal margin. Mean per-token log-likelihood of each story under the pretrained model at softmax temperature~${T=1}$ versus the measured full-model maximal margin, attained for the smallest~${\alpha}$ (largest~${N}$) in Fig.~\ref{fig:6}A (gray dots: 104 different stories).}\label{fig:9}
\end{figure}

The theory also describes the dependence of latent and functional connection norms on the worst relative bias (Fig.~\ref{fig:8}C--D). Because maximizing the margin fixes only a feasible region for each latent, not the latent itself, we draw its direction uniformly from the feasible directions and then minimize its norm, in both simulations and theory (Appendix~C). Both amplitudes increase with~${\Delta_\mu}$, and the target and rival connections approach opposite points on the unit sphere for sufficiently strong unfavorable biases. The theoretical curves agree with pair-model simulations and reproduce the trends observed in the full-model solutions (Fig.~\ref{fig:6}C--D). Thus the sparse pair construction captures both the enhanced retrieval margin and the heterogeneous strengths of the learned representations.

Repeated targets are accommodated by assigning positive biases to all but one occurrence of each target. Although each occurrence could impose its own constraint, this choice is consistent with the sparsity of negative biases in the LLM, and a repeated target then adds an association without adding a new constraint. Unique targets already predicted by the pretrained model are similarly inert. Both effects enter the solution through a straightforward rescaling of~${\alpha}$.

In the large-vocabulary regime~$\beta=M/N\gg1$, both the fixed-$\boldsymbol{\hat{x}}$ and trained-$\boldsymbol{\hat{x}}$ models have the same leading capacity scaling,
\begin{equation}
\label{eq:12}
\alpha\approx C_p\left(\kappa\right)\frac{\bar{n}R}{\log\beta},
\end{equation}
where~$C_p(\kappa)=O\left(1\right)$ is given explicitly in Appendix~B, and~${\bar{n}}$ is the ratio of the sequence length to the number of unique targets that the pretrained model does not predict. For example,~${\bar{n}=1}$ for a sequence of distinct targets, none of which is predicted. Thus, the weak logarithmic dependence on vocabulary size and linear dependence on~${R}$ in the LoRA-only model extend to this more general setting, suggesting it may hold for any choice of the pretrained relative biases.

\paragraph{Story predictability and the maximal margin.}
The pair model also accounts for how the predictability of a story affects its storage. Its maximal margin depends on a single parameter, $\kappa_{\max}=2/(1+\Delta_{\max})$ (Eq.~\eqref{eq:pair_kappa_max}), so the margin of each story can be matched by an effective maximal rival bias. Across~104 stories, the mean per-token log-likelihood under the pretrained model explains 60\% of the variance of the measured full-model maximal margin (Fig.~\ref{fig:9}). More predictable stories are therefore stored with larger margins, consistent with the role of prior knowledge in facilitating storage. In the pair model, these stories behave as if their strongest unfavorable biases were weaker.

\section{Efficiency}\label{sec:efficiency}

We next quantify the parameter efficiency~$e$ of the LoRA architecture. We define efficiency as the number of stored associations per trainable parameter. We note that the LoRA matrices contain
\begin{equation}
RN_0+RN
\end{equation}
trainable parameters, and realizing arbitrary latent representations requires~${N_0\geq P}$ (see below). Thus, the maximal efficiency is obtained for~${N_0=P}$, giving
\begin{equation}
\label{eq:16}
e=\frac{P}{RP+RN}=\frac{1}{R}\frac{\alpha}{\alpha+1}.
\end{equation}

\begin{figure}
\centering{}\includegraphics[width=\columnwidth]{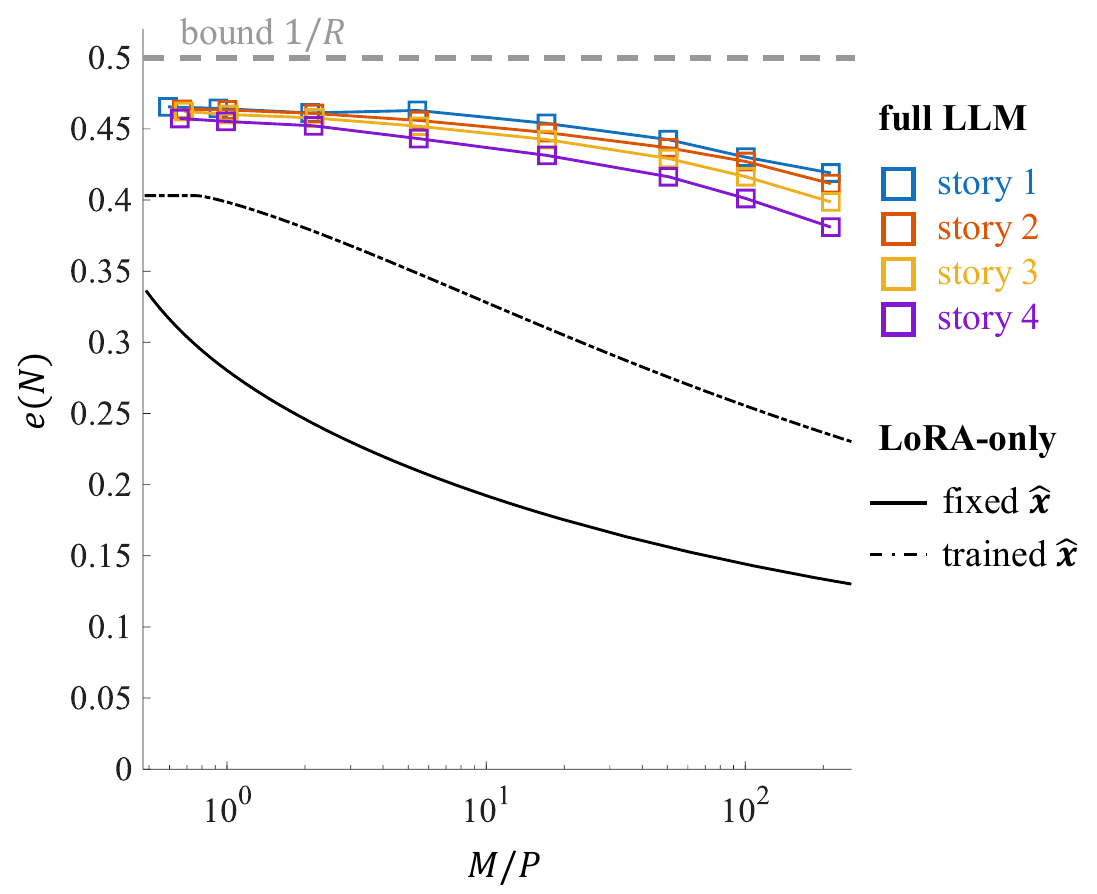}\caption{Efficiency, defined  as the number of associations per trainable parameter at capacity, versus the ratio~${M/P}$ of the vocabulary size to the sequence length, for~${R=2}$. Lines show the LoRA-only theory for fixed and trained latent directions, evaluated at the mean token multiplicity~${m=2.08}$ of the four stories by an exact rescaling of the distinct-target theory (Sec.~\ref{sec:LoRA-only}). Squares show simulations of the full LLM for four stories of length~${P=600}$, whose efficiency reaches ${\sim80\%}$ of the upper bound~${1/R}$. Each value of~${M/P}$ is obtained by subsampling~${M-P_{\text{unique}}}$ distractors, so that~$M$ ranges from each story's~$P_{\text{unique}}$ up to the full LLM vocabulary, ${M=128{,}256}$.}\label{fig:10}
\end{figure}

Although~$M$ does not appear explicitly in this expression, it enters through the capacity~$\alpha$, which depends on the vocabulary load~$\beta=M/N=\alpha(M/P)$. Figure~\ref{fig:10} shows the resulting efficiency as a function of~${M/P}$, the ratio of the vocabulary size to the sequence length. The LoRA-only curves are evaluated at the mean token multiplicity of the four stories, ${m=2.08}$. Training the latent directions~$\boldsymbol{\hat{x}}_{\mu}$ consistently improves the efficiency over holding them fixed, by up to a factor of~${1.8}$. Even for~$M/P\gg1$, the efficiency decreases only slowly, because the capacity falls only logarithmically with the vocabulary.

For the minimal~${N_0=P}$, each association adds an additional~${R}$ input connections used to create its latent representation~$\boldsymbol{x}_\mu$. When~${\alpha\ll1}$, these~${PR}$ input connections are a small fraction of the total LoRA parameters, so by Eq.~\eqref{eq:16} the efficiency is roughly proportional to the capacity:~${e\approx\alpha/R}$. When~${\alpha \gg 1}$, the input connections dominate, and the efficiency saturates at~${e\approx 1/R}$. In LLMs, as in episodic memory generally, the space of possible tokens or events is orders of magnitude larger than any single memory (${M/P\approx200}$ here). At such vocabularies, storage without prior knowledge is limited by capacity: the LoRA-only model reaches only~${\alpha\approx0.8}$, where the efficiency is about half its bound. The pretrained network increases the capacity by a factor of 4--6 and the efficiency by a factor of~1.7--1.9, to ${\sim80\%}$ of its upper bound (rightmost points in Fig.~\ref{fig:10}). Hence, in this regime, capacity drives efficiency.

In every case, the efficiency in Eq.~\eqref{eq:16} is~$O\left(1\right)$ in the feature layer dimension~$N$, in contrast to conventional associative memories. For example, a Hopfield network requires~$O\left(N^{2}\right)$ trainable parameters to store~$O\left(N\right)$ patterns, giving~$e_{\text{Hopfield}}=O\left(N^{-1}\right)$. Modern Hopfield networks similarly require $O\left(N\right)$ parameters to explicitly represent each stored pattern~\citep{krotov2016}. In our model, by contrast, each additional association is encoded through the low-rank weights while the output vocabulary and its readout remain fixed.

This high efficiency arises from two features of the architecture. First, retrieval is a winner-take-all decision: only the identity of the winning output neuron matters, so different output states sharing a winner are functionally interchangeable. This sharply relaxes the storage constraints compared to recurrent attractor networks, which must reconstruct the full output pattern. Second, the vocabulary is fixed, so the readout matrix need not be modified when a new sequence is stored. Together, these features allow the number of trainable parameters required per association to remain $O\left(1\right)$, providing a simple mechanism for parameter-efficient continual storage.

\paragraph{Realizing the latent representations.}
So far, we have characterized the storage capacity assuming that the latent representations~$\boldsymbol{x}_{\mu}$ can be trained independently. We now ask how these representations can be realized by the LoRA matrix~$V$. Let~${\Xi\in\mathbb{R}^{N_{0}\times P}}$ denote the matrix whose columns are the inputs~${\boldsymbol{\xi}\left(\boldsymbol{e}_{\mu}\right)}$ to the final MLP down projection. When these~${P}$ inputs are linearly independent, which requires~${N_0\geq P}$, the minimum-norm solution that realizes a learned set of latent representations~${X\in\mathbb{R}^{P\times R}}$ is
\begin{equation}
V=\Xi\left(\Xi^{T}\Xi\right)^{-1}X
\end{equation}
where~$X$ may contain duplicate latent representations corresponding to repeated target tokens.

\begin{figure}[t]
  \centering
  \includegraphics[width=\columnwidth]{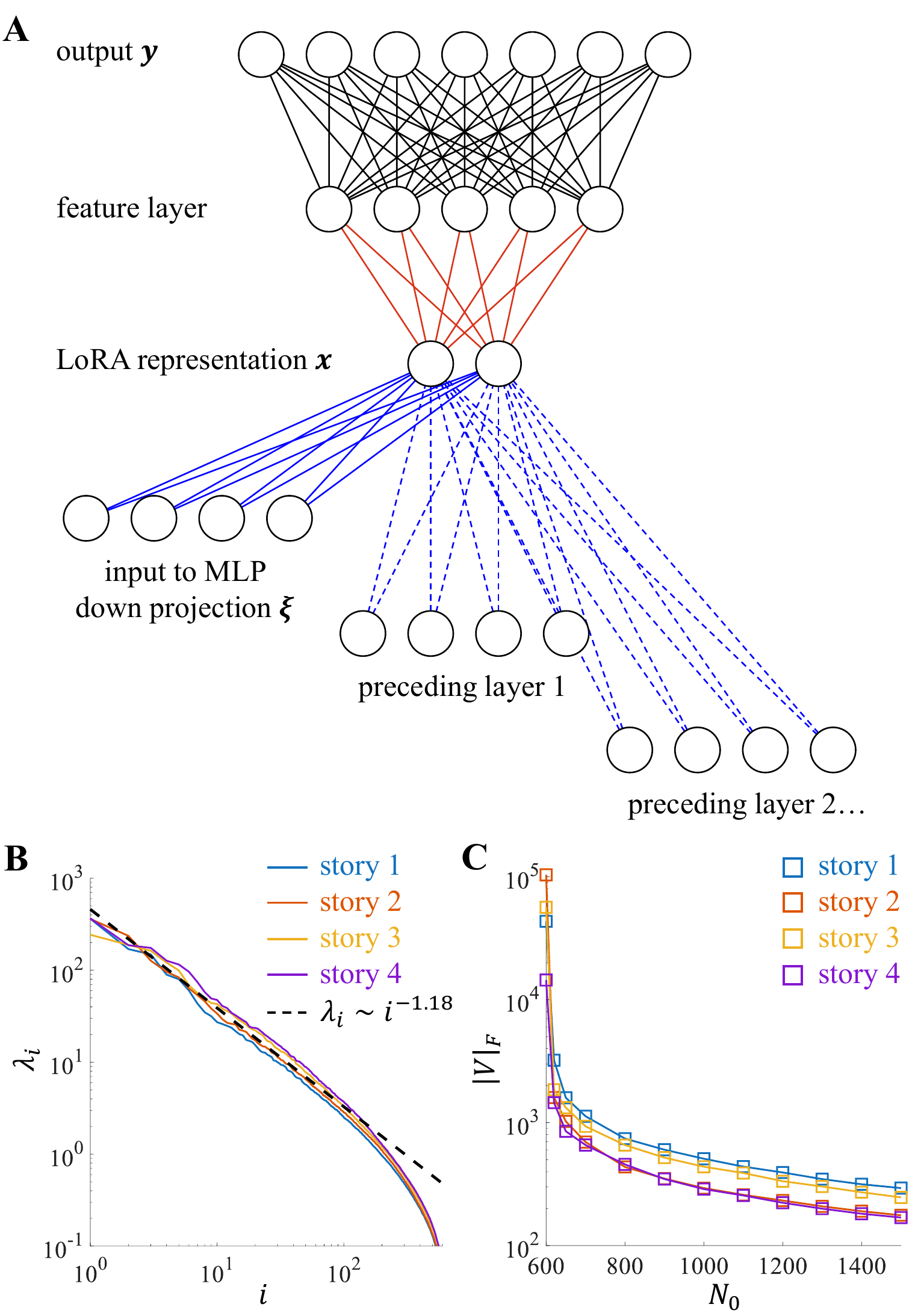}\caption{\textbf{A.} Previous LLM layers can be recruited to increase~$N_0$, provided the resulting inputs are linearly independent across positions. \textbf{B.} The eigenvalue spectrum of~${\Xi^\top\Xi}$ follows an approximate power law, with exponent~$1.18$, across multiple stories. \textbf{C.} Increasing~$N_0$ slightly beyond the minimum value~$P$ (here~${P=600}$) substantially reduces the Frobenius norm of the minimum-norm solution~$V$, which may limit interference with the pretrained model.}\label{fig:11}
\end{figure}

For sequences longer than the input width of the final MLP block, this requirement can still be met by adding input neurons. One possibility is to recruit neurons from earlier LLM layers and connect them to the feature layer through skip connections, effectively increasing $N_0$ to the number of activation units, which can reach~$10^6$--$10^7$ in most LLMs (Fig.~\ref{fig:11}A). For this construction to work, the additional activations must make the inputs at different positions linearly independent, and they must be deterministic functions of the input~$\boldsymbol{e}_{\mu}$, so that repeated presentations of the same input produce the same latent representation. The activations of the LLM satisfy both conditions. Since the total number of activations, in LLMs as in the brain, vastly exceeds the length of any single episode, this constraint is unlikely to limit storage in practice.

The inputs are strongly correlated: the eigenvalues of~${\Xi^\top\Xi}$, computed with all input neurons, approximately follow a power law with exponent~$1.18$ (Fig.~\ref{fig:11}B). To map such inputs onto the latent representations~${\boldsymbol{x}_\mu}$, $V$ must suppress these correlations by amplifying the weakest eigendirections, which inflates its norm~${\|V\|_F}$. The norm also depends on~${N_0}$: increasing~$N_0$ even slightly beyond its minimum value~$P$ substantially reduces it (Fig.~\ref{fig:11}C), which may limit interference of the LoRA update with the pretrained model's existing functionality.

\section{The role of the margin in recall}

\begin{figure}
\centering{}\includegraphics[width=\columnwidth]{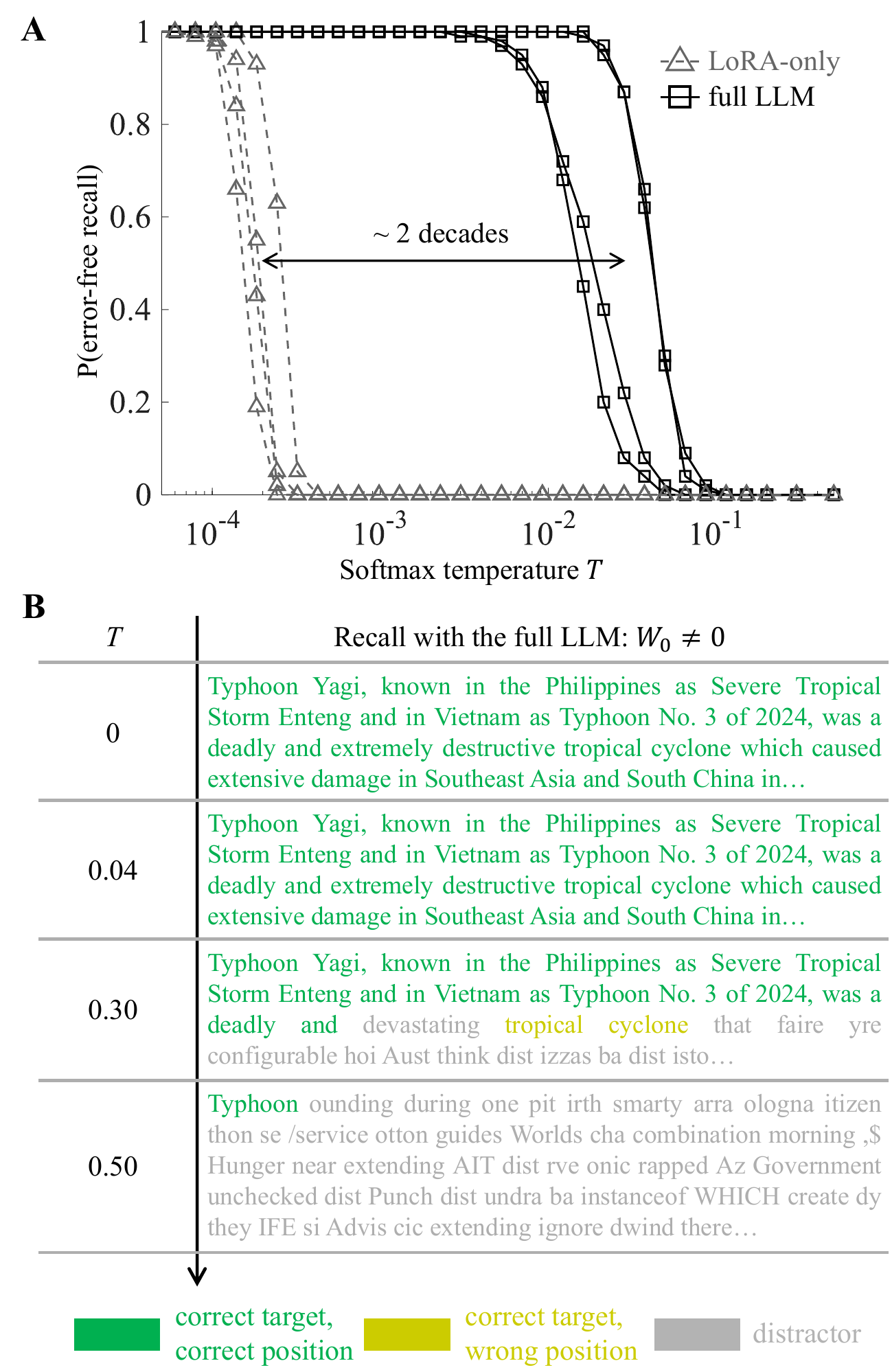}\caption{A finite margin makes recall robust to sampling noise. \textbf{A.} Probability that a whole story of length~${P=600}$ is recalled without error, versus~${T}$, for the LoRA-only model (gray triangles) and the full LLM (black squares), for the four stories in Fig.~\ref{fig:6}A. Each probability is estimated from 100 trials, and in both models the logit vector is normalized to~${|\boldsymbol{y}|=800}$ (see text). \textbf{B.} First~50 tokens generated by the fine-tuned full LLM for one story at increasing softmax temperatures~${T}$. Green: correct target in the correct position; yellow: correct target in the wrong position; gray: distractor. Simulations use~${R=2}$, ${N_0=1200}$, ${N=1000}$, and the full LLM vocabulary~${M=\text{128,256}}$.}\label{fig:12}
\end{figure}

So far we have treated retrieval as a deterministic winner-take-all. In an LLM, however, each token is sampled from the softmax distribution over the logits, ${p_n \propto e^{y_n/T}}$, so the correct event is selected only with high probability. Here we show that a finite margin is what allows recall to survive this sampling noise.

An episode is recalled correctly only if no error occurs at any of its~${P}$ positions, which requires a per-token error probability~${p_{\mathrm{err}}\lesssim1/P}$. Since~${p_{\mathrm{err}}}$ decays exponentially with the ratio of the logit gap between the target and its strongest competitor to~${T}$, recall succeeds below a threshold temperature~${T_c}$ and fails abruptly above it, and a larger margin raises~$T_c$. The full LLM, which has a finite margin, should therefore differ sharply from the LoRA-only model, whose margin vanishes as~${\kappa \sim P^{-2}}$ for~${R=2}$.

Because sampling depends on the logits through~${\boldsymbol{y}/T}$, comparing temperatures across models requires a common logit scale. In the LoRA-only model this scale is set by the norms~${|\boldsymbol{x}_\mu|}$, which do not affect the WTA decision. We choose them so that~${|\boldsymbol{y}|=800}$, the typical scale of the pretrained LLM's output (${|\boldsymbol{y}(\boldsymbol{e}_\mu)| = 810 \pm 150}$ across all~${\mu}$ and all stories analyzed here), and normalize the full LLM's logits to the same value at each position. In the full LLM, we set each~${|\boldsymbol{x}_\mu|}$ to the larger of~1 and the minimal value consistent with the constraints, so that no logit gap vanishes.

Figure~\ref{fig:12} confirms that the two models differ sharply. Taking~${T_c}$ as the temperature at which the probability of error-free recall falls to one half, we find across four stories~$T_c=0.02$--$0.04$ for the full LLM, but only~${T_c\sim2\times 10^{-4}}$ in the LoRA-only model, two orders of magnitude lower (Fig.~\ref{fig:12}A). Figure~\ref{fig:12}B illustrates how recall in the full LLM degrades as the temperature increases. At~${T=0.04}$, the first~50 tokens are reproduced exactly. At~${T=0.30}$, the opening of the story is reproduced, after which the text starts to deviate. At~${T=0.50}$, the text departs from the story immediately after its first word. The finite margin supplied by the pretrained network thus makes retrieval robust to the stochasticity of sampling.

\section{Discussion}

We developed a statistical-mechanical theory of episodic memory storage by low-rank adaptation of a pretrained network, motivated by complementary learning systems in the brain and by a recent architecture that assigns each episode its own adapter in an LLM~\citep{pan2026}. The key simplification afforded by the LLM architecture is that sequential memory storage becomes a supervised classification problem over a finite, fixed vocabulary. Without the pretrained connectivity, the solution acquires an emergent spherical geometry, with target connections converging onto the unit sphere along their latent directions. The nonconvexity of the storage constraints breaks replica symmetry, and the one-step solution agrees closely with simulations. Two features emerge. First, storage by a low-rank adapter is efficient, requiring only~${O(1)}$ trainable parameters per association. This is made possible by WTA retrieval, which only needs the correct event to win,  without reconstructing a stored pattern. Second, capacity decreases only as the inverse logarithm of the vocabulary size, so a small adapter can store episodes drawn from an enormous space of possible events.

Restoring the pretrained connectivity changes the problem qualitatively. Before fine-tuning, the base model already predicts a substantial fraction of the associations, and for most of the rest it favors the correct output over all but a few competitors. As a result, the retrieval margin becomes finite instead of vanishing, and the capacity increases. The pretrained relative biases also break the homogeneity of the constraints: functional connections leave the unit sphere, with norms that grow with the strength of the competing bias, and latent representation norms grow with the surprise of each association's next event under the pretrained network. A solvable model of sparse pretrained relative biases, which retains for each target a single competitor that the pretrained network wrongly favors, reproduces each of these features: the finite margin, the increased capacity, the departure of functional connections from the unit sphere, and the dependence of connection strengths and representation norms on surprise.

A central analytical step is the reduction of the ${P(M-1)=O(N^2)}$ inequalities in Eq.~\eqref{eq:3} to a number of effective constraints linear in~$N$. In the LoRA-only model, the emergent spherical geometry replaces the pairwise competitions by alignment or norm constraints on the target connections (Eqs.~\eqref{eq:10} and~\eqref{eq:7}), together with norm bounds on the distractor connections. In the pair bias model, the assumed structure of the pretrained relative biases reduces the~$M-1$ comparisons for each target to the single constraint in Eq.~\eqref{eq:15}. Including the~$O(M)$ bounds on functional connection norms, the number of constraints remains~$O(N)$ at fixed rank and loads. This reduction makes the replica mean-field analysis tractable.

Our results suggest a simple physical interpretation of learning by adapting a pretrained computational substrate. Rather than constructing an entirely new memory, the adapter exploits the pretrained substrate to resolve most of the classification constraints, leaving only a sparse set to be corrected. This sparsity increases the storage capacity and, more dramatically, the robustness of retrieval. Computational neuroscience models have similarly proposed that the hippocampus stores what the neocortex cannot predict, making efficient use of limited hippocampal storage~\citep{spens2024,fayyaz2022}. Our theory makes this idea quantitative, deriving prediction-error coding from optimal storage with minimal latent norms instead of assuming it. Consistent with this, hippocampal responses to unexpected outcomes are larger when the violated prediction is stronger~\citep{long2016,chen2015}.

We have focused here on memory for a single episode, and our optimization constrains the network only on the stored contexts. Once sampling departs from the stored sequence, the fine-tuned model produces incoherent text (Fig.~\ref{fig:12}B), showing that the adapter can interfere with the pretrained predictions on other contexts. Important directions for future work are therefore to determine how encoding an episode affects the performance of the combined base model and LoRA adapter on other tasks, and how multiple episodes can be stored without catastrophic forgetting. In separate work, Pan et al.~\citep{pan2026} proposed a gating mechanism that selects among episode-specific adapters to suppress such interference. Our analysis suggests a complementary possibility: episodic learning can be viewed as a form of sparse encoding, because the corrections needed to overcome unfavorable relative biases concern only a small subset of functional connections to targets and their unfavorable competitors, while the remaining connections need only satisfy a norm bound. It will be interesting to test whether this sparsity can be exploited to mitigate interference, even without gating.


Our framework also suggests a simple analogue of systems consolidation, in which memories that initially depend on the hippocampus are gradually incorporated into the neocortex~\citep{mcclelland1995,spens2024}. Merging an adapter into the base weights, $W_0\to W_0+UV^\top$, incorporates the learned episode into the base connectivity while preserving the predictions of the combined model, including any interference on other contexts. Recall of that episode then no longer requires a separate adapter. Whether such consolidation can facilitate the learning of related episodes by providing more favorable relative biases remains an open question.

A further limitation is that our pair model retains only one rival per association, whereas in a real LLM each target can have several unfavorable competitors, often shared with other targets. Extending the theory to this more general case remains an open problem.

\begin{acknowledgments}
We thank Ely Hahami for help with data acquisition. This research was supported by the Kempner Institute for the Study of Natural and Artificial Intelligence at Harvard University, the Office of Naval Research (Grant No.~N00014-23-1-2051), the Gatsby Charitable Foundation, and an Amazon Research Award (Spring 2025). H.St.\ was supported by a National Defense Science and Engineering Graduate (NDSEG) Fellowship, and this material is based upon work supported by the Air Force Office of Scientific Research under award number FA9550-23-F-0014 in the amount of \$140,900. D.D.L.\ was supported by the Institute of Information \& communications Technology Planning \& Evaluation (IITP) grant funded by the Korean government (MSIT) (No.~RS-2024-00457882, National AI Research Lab Project).
\end{acknowledgments}

\appendix
\section{Effective storage constraints and the maximal margin}\label{sec:appendix_A}

In this and the following appendices, we will first derive results for the pair model, and then recover the corresponding results for the LoRA-only model as a limiting case. We first derive the reduction of the WTA constraints used in the pair model. For target~$\mu$, the constraints are
\begin{equation}
\label{eq:A2}
\boldsymbol{\hat{x}}_{\mu}\cdot\left(\boldsymbol{h}_{T\left(\mu\right)}-\boldsymbol{h}_{n}\right)+\left|\boldsymbol{x}_{\mu}\right|^{-1}\delta^{n}_{\mu}\geq\kappa
\end{equation}
where~$\delta^{n}_{\mu}$ is~$-\Delta_{\mu}$ if~$n=R\left(\mu\right)$ and is~$1$ otherwise. To reduce these constraints in the thermodynamic limit, we make the assumption (confirmed in simulations) that the unit sphere~${S^{R-1}}$ is densely covered by some of the functional connections~$\boldsymbol{h}$. Then for fixed~$\mu$, we can replace the group of~${M-2}$ constraints for all other competitors,~$n\neq R\left(\mu\right)$, with a ``worst-case'' constraint for which~$\boldsymbol{\hat{x}}_{\mu}\cdot\boldsymbol{h}_{n}\approx1$. This constraint is
\begin{equation}
\boldsymbol{\hat{x}}_{\mu}\cdot\boldsymbol{h}_{T\left(\mu\right)}-1+\left|\boldsymbol{x}_{\mu}\right|^{-1}\geq\kappa\label{eq:13}
\end{equation}
The remaining WTA constraints are for~$n=R\left(\mu\right)$ and have the form
\begin{equation}
\boldsymbol{\hat{x}}_{\mu}\cdot\left(\boldsymbol{h}_{T\left(\mu\right)}-\boldsymbol{h}_{R\left(\mu\right)}\right)-\left|\boldsymbol{x}_{\mu}\right|^{-1}\Delta_{\mu}\geq\kappa\label{eq:14}
\end{equation}
Combining inequalities~\eqref{eq:13} and~\eqref{eq:14} and eliminating~$\left|\boldsymbol{x}_{\mu}\right|^{-1}$ yields Eq.~\eqref{eq:15}.

In the case where~${\boldsymbol{\hat{x}}_{\mu}}$ is trained, this constraint can be simplified even further: there exists~$\boldsymbol{\hat{x}}_{\mu}$ satisfying~\eqref{eq:15} if and only if
\begin{equation}
\label{eq:A4}
\left|\left(1+\Delta_\mu\right)\boldsymbol{h}_{T(\mu)}-\boldsymbol{h}_{R(\mu)}\right|\geq \Delta_\mu\kappa+\Delta_\mu+\kappa,
\end{equation}
a nonconvex cylindrical constraint in the $2R$-dimensional space of~${\left(\boldsymbol{h}_{T(\mu)},\boldsymbol{h}_{R(\mu)}\right)}$.

The LoRA-only model is recovered by taking~${\Delta_{\mu}\rightarrow\infty}$ for all~$\mu$. Dividing Eq.~\eqref{eq:15} by $\Delta_\mu$ and taking this limit gives $\hat{\boldsymbol x}_\mu\cdot\boldsymbol h_{T(\mu)}\ge1+\kappa$, which together with $|\boldsymbol h_{T(\mu)}|\le1$ yields Eq.~\eqref{eq:10}:
\begin{equation}
\boldsymbol{h}_{T\left(\mu\right)}\rightarrow\boldsymbol{\hat{x}}_{\mu},\quad\kappa\rightarrow0 .
\end{equation}
On the other hand, if~${\boldsymbol{\hat{x}}}$ is trained, then taking~${\Delta_\mu\rightarrow\infty}$ in Eq.~\eqref{eq:A4} yields Eq.~\eqref{eq:7}:
\begin{equation}
|\boldsymbol{h}_{T\left(\mu\right)}|\rightarrow1,\quad\kappa\rightarrow0.
\end{equation}

We now derive the scaling of the maximal margin in the LoRA-only model, Eq.~\eqref{eq:4}, by a packing argument. Let~${\theta(P)}$ be the largest minimum angular separation attainable by~${P}$ points on~${S^{R-1}}$. For an optimal configuration, the spherical caps of angular radius~${\theta(P)/2}$ must be disjoint, and each cap has area~${\sim\theta^{R-1}}$, so~${P\theta(P)^{R-1} \lesssim 1}$. Conversely, let~${P(\theta)}$ be the largest number of points separated by at least~${\theta}$. In such a configuration, the caps of radius~${\theta}$ must cover the sphere, since otherwise a further point could be added, so~${P(\theta)\theta^{R-1} \gtrsim 1}$. Setting~${\theta\sim P^{-1/(R-1)}}$ gives~${P(\theta) \gtrsim P}$, so a configuration of~${P}$ points with this separation exists, and therefore~${\theta(P) \gtrsim P^{-1/(R-1)}}$. Combining this with the first inequality yields~${\theta(P) \sim P^{-1/(R-1)}}$, and thus, since~${\boldsymbol{h}_\mu = \boldsymbol{\hat{x}}_\mu}$ and the binding constraints are those between pairs of targets,
\begin{equation}
    \kappa_{\max}(P) = 1 -\cos\theta(P) \sim \theta(P)^2 \sim P^{-2/(R-1)}.
\end{equation}
The coefficient is set by the optimal packing density in~${R-1}$ dimensions.

\section{Fixed $\boldsymbol{\hat{x}}$ replica calculation}\label{sec:appendix_B}

For fixed latent directions~$\boldsymbol{\hat{x}}_{\mu}$, the effective pair model constraints \eqref{eq:15} are convex in the functional connections. We therefore use the replica-symmetric (RS) ansatz to compute the volume of solutions with margin~${\kappa}$, given by
\begin{equation}
\begin{split}
Z=&\int dU\prod^{P}_{\mu=1}\Theta\Big(\boldsymbol{\hat{x}}_{\mu} \cdot \left( \left( 1 + \Delta_\mu \right) \boldsymbol{h}_{T\left(\mu\right)} - \boldsymbol{h}_{R\left(\mu\right)} \right)\\
&- \Delta_{\mu}\kappa - \Delta_{\mu} - \kappa\Big),
\end{split}
\end{equation}
where~${\boldsymbol{h}_n = U^\top \boldsymbol{a}_n}$ and is subject to the normalization~$\left|\boldsymbol{h}_{n}\right|\leq1$. The quenched free energy is obtained from
\begin{equation}
\label{eq:A5}
f=-\lim_{N\rightarrow\infty}\frac{1}{N}\lim_{n\rightarrow0}\frac{\left[Z^{n}\right]-1}{n},
\end{equation}
where~$\left[\cdot\right]$ denotes the quenched average over the Gaussian readout matrix $A$, the i.i.d. biases~$\Delta_{\mu}\sim p$, and the latent directions~$\boldsymbol{\hat{x}}_{\mu}$. For the latter, we assume only that the~$\left\{\boldsymbol{\hat{x}}_{\mu}\right\}$ are equidistributed on the sphere $S^{R-1}$ and are independent of their respective readout vectors~${\boldsymbol{a}_\mu}$ and biases~$\Delta_{\mu}$, a condition met both by i.i.d. uniform directions and by a random permutation of the optimal packing arrangement of the~$\boldsymbol{\hat{x}}_{\mu}$ discussed in Sec.~\ref{sec:LoRA-only} and Appendix~A.

We introduce the overlap matrix
\begin{equation}
\label{eq:A7}
q^{rs}_{ab}=\frac{1}{N}\boldsymbol{u}^{a}_{r}\cdot\boldsymbol{u}^{b}_{s},
\end{equation}
where~$a$ and~$b$ are replica indices and~$\boldsymbol{u}^{a}_{r},\,1\leq r\leq R$ denotes the~$r^{\text{th}}$ column of the LoRA matrix~$U$. After taking the quenched average, the replicated free energy separates into an entropic part, an energetic part from the~${P}$ target-rival pairs, and another energetic part from the remaining~$M-2P$ distractors that are not selected as rivals:
\begin{equation}
\label{eq:A6}
f\left(Q\right)=G_{S}\left(Q\right)+\alpha G^{1}_{E}\left(Q\right)+\left(\beta-2\alpha\right)G^{2}_{E}\left(Q\right).
\end{equation}
Imposing the replica-symmetric ansatz~$q^{rs}_{ab}=\delta_{rs}q_{ab}$, where
\begin{equation}
\label{eq:A8}
q_{ab}=\begin{cases}
q_{d}, & a=b,\\
q, & a\neq b,
\end{cases}
\end{equation}
and taking the~$n\rightarrow0$ limit, the three contributions reduce to
\begin{equation}
\begin{split}
&G_{S}\left(q_{d},q\right) = -\frac{R}{2}\left(\frac{q}{q_{d}-q}+\log\left(q_{d}-q\right)+\log\left(2\pi e\right)\right)\\
&G^{1}_{E}\left(q_{d},q\right) = R\left(\log\left(q_{d}-q\right)+\log\left(2\pi\right)\right)\\
 &\quad -\Bigg\langle \log\int d\boldsymbol{h}\,d\boldsymbol{g}\,K_q(\boldsymbol{h},\boldsymbol{z})\,K_q(\boldsymbol{g},\boldsymbol{y})\\
 &\quad\cdot\Theta\left((1+\Delta)h_{1}-g_{1}-\Delta\kappa-\Delta-\kappa\right)\Bigg\rangle _{\boldsymbol{z},\boldsymbol{y},\Delta}\\
&G^{2}_{E}\left(q_{d},q\right) = \frac{R}{2}\left(\log\left(q_{d}-q\right)+\log\left(2\pi\right)\right)\\
&\quad -\left\langle \log\int d\boldsymbol{h}\,K_q(\boldsymbol{h},\boldsymbol{z})\right\rangle _{\boldsymbol{z}}
\end{split}
\end{equation}
where the integrals over~$\boldsymbol{h}$ and~$\boldsymbol{g}$ are in the unit ball and~${\left(\boldsymbol{h},\boldsymbol{g}\right)}$ represents an effective target-rival pair~${\left(\boldsymbol{h}_{T\left(\mu\right)},\boldsymbol{h}_{R\left(\mu\right)}\right)}$, $\boldsymbol{z}$ and~$\boldsymbol{y}$ are $R$-dimensional Gaussian vectors, ${\Delta}$ has probability distribution~${p(\Delta)}$, and
\begin{equation}\label{eq:A13}
    K_q(\boldsymbol{h},\boldsymbol{z}) = \exp\left(-\frac{\left|\boldsymbol{h}-\sqrt{q}\boldsymbol{z}\right|^{2}}{2\left(q_{d}-q\right)}\right) .
\end{equation}

At capacity, the RS saddle approaches~$q_d-q\rightarrow0^+$. We therefore expand the integrals over~$\boldsymbol{h}$ and~$\boldsymbol{g}$ above by Laplace's method for~$q_d-q\ll1$, obtaining an asymptotic expression for the free energy of the form
\begin{equation}
f\left(q_{d},q\right)=\frac{g\left(q\right)}{q_{d}-q}+O\left(\log\left(q_{d}-q\right)\right).
\end{equation}
The coefficient~$g\left(q\right)$ of the leading~$1/\left(q_{d}-q\right)$ singularity is
\begin{equation}
\begin{split}
g\left(q\right) =& -\frac{1}{2}Rq+\frac{1}{2}\alpha\left\langle I\left(\boldsymbol{z},\boldsymbol{y},\Delta,q,\kappa\right)\right\rangle _{\boldsymbol{z},\boldsymbol{y},\Delta}\\
& +\frac{1}{2}\left(\beta-2\alpha\right)\left\langle \left[\sqrt{q}\left|\boldsymbol{z}\right|-1\right]^{2}_{+}\right\rangle _{\boldsymbol{z}},
\end{split}
\end{equation}
where
\begin{equation}
\begin{split}
&I(\boldsymbol z,\boldsymbol y,\Delta,q,\kappa)=\min_{h_1,g_1\in[-1,1]}\Bigg\{\big(h_1-\sqrt q\,z_1\big)^2\\
&\quad+\big(g_1-\sqrt q\,y_1\big)^2+\Big[\sqrt q\,|\boldsymbol z_{2\to R}|-\sqrt{1-h_1^2}\Big]_+^2\\
&\quad+\Big[\sqrt q\,|\boldsymbol y_{2\to R}|-\sqrt{1-g_1^2}\Big]_+^2\Bigg\}\\
&\quad\text{subject to}\quad (1+\Delta)h_1-g_1\geq\Delta\kappa+\Delta+\kappa .
\end{split}
\end{equation}
Here,~${\boldsymbol{z}_{2\rightarrow R} = \left[ z_2,\dots,z_R \right]}$ and~${\boldsymbol{y}_{2\rightarrow R}}$ is defined similarly. At capacity,~$g\left(q\right)$ must vanish at a stationary point. Thus the capacity conditions are
\begin{equation}
\label{eq:11}
g\left(q\right)=0,\quad\frac{\partial g}{\partial q}=0.
\end{equation}

\subsection{The large-vocabulary limit: $\beta\gg1$}

For the mean-field equations to have a solution in this limit, we require that the third term in~$g\left(q\right)$ have the same order as the first two terms. This requires
\begin{equation}
\left\langle \left[\sqrt{q}\left|\boldsymbol{z}\right|-1\right]^{2}_{+}\right\rangle _{\boldsymbol{z}}\sim1/\beta
\end{equation}
from which we obtain
\begin{equation}
q \approx \frac{1}{2\log\beta} + O\left( (\log\beta)^{-2}
\right)
\end{equation}
Now, we calculate the free energy to leading order in~$q$ in the regime~$q\ll1$. The leading order of~$I\left(\boldsymbol{z},\boldsymbol{y},\Delta,q,\kappa\right)$ is independent of~${\boldsymbol{z}}$ and~${\boldsymbol{y}}$ and given by
\begin{equation}
\begin{split}
I_0 \left( \Delta,\kappa \right)
=& \min_{h_{1},g_{1}\in[-1,1]}\Bigg\{h_{1}^{2}+g_{1}^{2}\quad\text{subject to}\quad\\
 &(1+\Delta)h_{1}-g_{1}-\Delta\kappa-\Delta-\kappa\geq0\Bigg\}
\end{split}
\end{equation}
The solution to the minimization is piecewise, depending on whether the optimal~${h_1}$ lies in the interior of the interval~${[-1,1]}$ or saturates the boundary~${h_1=1}$. For~${\kappa\leq2/(1+\Delta)}$, beyond which the constraint cannot be satisfied, the final result is
\begin{align}
\label{eq:A9}
I_0(\Delta,\kappa) &=\begin{cases}
\dfrac{\left(\Delta\kappa+\Delta+\kappa\right)^{2}}{\left(1+\Delta\right)^{2}+1}, & \kappa<\dfrac{2+\Delta}{\left(1+\Delta\right)^{2}},\\
1+\left(\Delta\kappa+\kappa-1\right)^{2}, & \text{otherwise}.
\end{cases}
\end{align}
The resulting leading free energy is
\begin{equation}
\label{eq:A14}
\begin{split}
g(q_1) \approx &-\frac{R}{2}q_1 + \frac{1}{2}\alpha \left\langle I_0\left(\Delta,\kappa\right)\right\rangle_{\Delta}\\
&+\beta\frac{|S^{R-1}|}{\left(2\pi\right)^{R/2}}e^{-\frac{1}{2q_1}}q_1^{3-R/2} ,
\end{split}
\end{equation}
where~${|S^{R-1}|}$ denotes the surface area of the unit sphere~${S^{R-1}}$. Solving the capacity conditions for this free energy yields
\begin{equation}
\alpha \approx \frac{1}{2\left\langle I_0\left(\Delta,\kappa\right)\right\rangle _{\Delta}}\frac{R}{\log\beta}
\end{equation}
To get Eq.~\eqref{eq:12}, we define
\begin{equation}
C_p(\kappa) = \frac{1}{2\left\langle I_0\left(\Delta,\kappa\right)\right\rangle _{\Delta}}
\end{equation}

\subsection{The LoRA-only limit: $\Delta\rightarrow\infty$}

The LoRA-only model is recovered by taking~${\Delta\rightarrow\infty}$, which forces~${\kappa\rightarrow0}$. In this limit we can write down a closed-form expression for~${I\left(\boldsymbol{z},\boldsymbol{y},\Delta,q,\kappa=0\right)}$. The constraint is met if and only if~${h_1=1}$, resulting in
\begin{align}
I\left(\boldsymbol{z},\boldsymbol{y},\Delta,q,0\right) & \rightarrow 1 - 2\sqrt{q}z_1 + q|\boldsymbol{z}|^2 + \left[ \sqrt{q} |\boldsymbol{y}| - 1 \right]_+^2
\end{align}
Thus
\begin{align}
\left\langle I\left(\boldsymbol{z},\boldsymbol{y},\Delta,q,0\right) \right\rangle_{\boldsymbol{z},\boldsymbol{y}} \rightarrow 1 + Rq + \left\langle \left[ \sqrt{q} |\boldsymbol{y}| - 1 \right]_+^2 \right\rangle_{\boldsymbol{y}}
\end{align}
Then~$g\left(q\right)$ becomes
\begin{equation}
\begin{split}
g\left(q\right) =& -\frac{1}{2}Rq+\frac{1}{2}\alpha\left(1+Rq\right)\\
&+\frac{1}{2}\left(\beta-\alpha\right)\left\langle \left[\sqrt{q}\left|\boldsymbol{z}\right|-1\right]^{2}_{+}\right\rangle _{\boldsymbol{z}}
\end{split}
\end{equation}
The resulting capacity conditions reduce to Eqs.~(\ref{eq:1}-\ref{eq:2}) of the main text.

\section{Trained $\boldsymbol{\hat{x}}$ replica calculation}\label{sec:appendix_C}

Here, we consider the case where $\boldsymbol{\hat{x}}_{\mu}$ is trained. As before, we compute the volume of solutions satisfying the storage constraints with margin $\kappa$, using the trained~${\boldsymbol{\hat{x}}}$ constraints~\eqref{eq:A4}:
\begin{equation}
\begin{split}
\label{eq:A10}
Z=&\int dU\prod^{P}_{\mu=1}\Theta\Big(\left|\left(1+\Delta_\mu\right)\boldsymbol{h}_{T(\mu)}-\boldsymbol{h}_{R(\mu)}\right|\\
&-\Delta_\mu\kappa-\Delta_\mu-\kappa\Big),
\end{split}
\end{equation}
The quenched average of~${Z}$ is taken over~${A}$ and~${\Delta_\mu}$. The quenched free energy~${f}$, the overlap matrix~${q^{rs}_{ab}}$, and the decomposition of~${f}$ into~${G_S}$, ${G_E^1}$, and~${G_E^2}$, are all defined as before~(\ref{eq:A5},\ref{eq:A7},\ref{eq:A6}).

We start by making the replica symmetric ansatz~\eqref{eq:A8}. In this case, taking the~${n\rightarrow0}$ limit, the three contributions reduce to
\begin{equation}
\begin{split}
&G_{S}\left(q_{d},q\right) =-\frac{R}{2}\left(\frac{q}{q_{d}-q}+\log\left(q_{d}-q\right)+\log\left(2\pi e\right)\right)\\
&G_E^1(q_d,q)=R\left(\log(q_d-q)+\log 2\pi\right)\\
&\quad-\Big\langle\log\!\int d\boldsymbol h\,d\boldsymbol g\,K_q(\boldsymbol h,\boldsymbol z)\,K_q(\boldsymbol g,\boldsymbol y)\,\\
&\quad\cdot\Theta\big(|(1+\Delta)\boldsymbol h-\boldsymbol g|-\Delta\kappa-\Delta-\kappa\big)\Big\rangle_{\boldsymbol z,\boldsymbol y,\Delta}\\
&G^{2}_{E}\left(q_{d},q\right) =\frac{R}{2} \left(\log\left(q_{d}-q\right)+\log\left(2\pi\right)\right)\\
&\quad-\left\langle \log\int d\boldsymbol{h}K_q(\boldsymbol h,\boldsymbol z)\right\rangle _{\boldsymbol{z}}
\end{split}
\end{equation}
where~${K_q(\boldsymbol h,\boldsymbol z)}$ is given by Eq.~\eqref{eq:A13}. Taking $q\rightarrow q_{d}$ yields
\begin{equation}
\begin{split}
g\left(q\right) =& -\frac{1}{2}Rq+\frac{1}{2}\alpha\left\langle I\left(\boldsymbol{z},\boldsymbol{y},\Delta,q,\kappa\right) \right\rangle_{\boldsymbol{z},\boldsymbol{y},\Delta} \\
&+\frac{1}{2}\left(\beta-2\alpha\right) \left\langle \left[\sqrt{q}\left|\boldsymbol{z}\right|-1\right]_+^{2} \right\rangle_{\boldsymbol{z}}
\end{split}
\end{equation}
where
\begin{equation}
\begin{split}
\label{eq:A3}
&I\left(\boldsymbol{z},\boldsymbol{y},\Delta,q,\kappa\right) =\min_{\boldsymbol{h},\boldsymbol{g}\in B^R}\Bigg\{ \left|\boldsymbol{h}-\sqrt{q}\boldsymbol{z}\right|^{2}+\left|\boldsymbol{g}-\sqrt{q}\boldsymbol{y}\right|^{2}\\
&\quad\text{subject to}\quad
\left|\left(1+\Delta\right)\boldsymbol{h}-\boldsymbol{g}\right|-\Delta\kappa-\Delta-\kappa\geq0\Bigg\}.
\end{split}
\end{equation}
Here,~${B^R}$ denotes the ${R}$-dimensional unit ball. The capacity conditions are again
\begin{equation}
g\left(q\right)=0,\quad\frac{\partial g}{\partial q}=0.
\end{equation}

However, we find that the RS solution does not agree with simulations. Hence, we replace the RS ansatz for~${q_{ab}}$ with the 1-step replica symmetry breaking ansatz,
\begin{equation}
q_{ab}=\begin{cases}
q_{d}, & a=b\\
q_{1}, & a\neq b,\Big\lceil\frac{a}{m}\Big\rceil=\Big\lceil\frac{b}{m}\Big\rceil\\
q_{0}, & \text{otherwise}
\end{cases}
\end{equation}
We take~${q_0=0}$, as suggested by the global rotational symmetry of the problem: if~${U}$ is a solution, so is~${UO}$ for any orthogonal~${O\in\mathbb{R}^{R\times R}}$, and in particular~${-U}$. The three contributions to the free energy are
\begin{equation}
\begin{split}
&G_{S}\left(q_{d},q_{1}\right)=-\frac{R}{2}\left(\Lambda_m +\log\left(2\pi e\right)\right)\\
&G^1_{E}\left(q_{d},q_{1}\right) =R\left(\Lambda_m+\log(2\pi)\right) - \frac{1}{m}\Bigg\langle\log \Bigg\langle \Bigg(\int d\boldsymbol{h}\,d\boldsymbol{g}\\
 &\quad\exp\Bigg[\frac{1}{q_{d}-q_{1}}\Bigg(-\frac{1}{2}\left(\left|\boldsymbol{h}\right|^{2}+\left|\boldsymbol{g}\right|^{2}\right)\\
 &\quad+\sqrt{\tilde{q}}\left(\boldsymbol{z}\cdot\boldsymbol{h}+\boldsymbol{y}\cdot\boldsymbol{g}\right)\Bigg)\Bigg]\\
 &\quad\cdot \Theta\left(\left|\left(1+\Delta\right)\boldsymbol{h}-\boldsymbol{g}\right|-\Delta\kappa-\Delta-\kappa\right)\Bigg)^{m} \Bigg\rangle_{\boldsymbol{z},\boldsymbol{y}}\Bigg\rangle_{\Delta}\\
&G^2_{E}\left(q_{d},q_{1}\right) =\frac{R}{2}\left(\Lambda_m+\log(2\pi)\right)\\
&\quad-\frac{1}{m}\log \Bigg\langle \Bigg(\int d\boldsymbol{h}\exp\Bigg[\frac{1}{q_{d}-q_{1}}\Big(-\frac{1}{2}\left|\boldsymbol{h}\right|^{2}\\
&\quad+\sqrt{\tilde{q}}\boldsymbol{z}\cdot\boldsymbol{h}\Big)\Bigg]\Bigg)^{m} \Bigg\rangle_{\boldsymbol{z}}
\end{split}
\end{equation}
where
\begin{equation}
    \Lambda_m = \log\left(q_{d}-q_{1}\right)+\frac{1}{m}\log\left(1+\frac{mq_{1}}{q_{d}-q_{1}}\right)
\end{equation}
and
\begin{equation}
    \tilde{q} = \frac{q_1}{1+mq_1/(q_d-q_1)}
\end{equation}
Next, we will take ${q_1\rightarrow q_d}$ and ${m\rightarrow0}$ such that ${\gamma=(q_d-q_1)/m}$ remains constant. This yields
\begin{equation}
\begin{split}
&g\left(q_1,\gamma\right) =-\frac{R}{2}\gamma\log\left(1+q_{1}/\gamma\right)\\
&\quad-\alpha\gamma\left\langle\log\left\langle\exp\left(-\frac{1}{2\gamma}I\left(\boldsymbol{z},\boldsymbol{y},\Delta,q_1,\kappa\right)\right)\right\rangle_{\boldsymbol{z},\boldsymbol{y}}\right\rangle_{\Delta}\\
&\quad-\left(\beta-2\alpha\right)\gamma\log \left\langle \exp\left(-\frac{1}{2\gamma}\left[\sqrt{q_{1}}\left|\boldsymbol{z}\right|-1\right]^{2}_{+}\right)\right\rangle_{\boldsymbol{z}}
\end{split}
\end{equation}
where~$I\left(\boldsymbol{z},\boldsymbol{y},\Delta,q_1,\kappa\right)$ is given by Eq.~\eqref{eq:A3}. The capacity conditions are
\begin{equation}
\label{eq:A1}
g\left(q_{1},\gamma\right)=0,\quad\frac{\partial g}{\partial q_{1}}=0,\quad\frac{\partial g}{\partial\gamma}=0.
\end{equation}

\subsection{Distributions of functional connections and latent norms}

To obtain the distribution of target functional connections, we fix the rival bias of one pair, ${\Delta_1=\Delta}$, and insert a factor of~${\delta(\boldsymbol{h}_{T(1)}-\boldsymbol{h})}$ for a single replica in Eq.~\eqref{eq:A10}. Repeating the calculation of the preceding section, in the limit~${q_1\rightarrow q_d}$, ${m\rightarrow0}$ at fixed~${\gamma}$, each state collapses onto the minimizer~${\left(\boldsymbol{h}^*,\boldsymbol{g}^*\right)}$ of Eq.~\eqref{eq:A3} and is weighted by~${w=e^{-I(\boldsymbol{z},\boldsymbol{y},\Delta,q_1,\kappa)/2\gamma}}$, giving
\begin{equation}
\label{eq:A11}
    P(\boldsymbol{h} \mid \Delta) = \frac{\langle w\, \delta \left( \boldsymbol{h}^* \left( \boldsymbol{z},\boldsymbol{y} \right) - \boldsymbol{h} \right)\rangle_{\boldsymbol{z},\boldsymbol{y}}}{\langle w \rangle_{\boldsymbol{z},\boldsymbol{y}}} ,
\end{equation}
where~${q_1}$ and~${\gamma}$ solve the capacity conditions~\eqref{eq:A1}. The distribution of rival connections follows by replacing~${\boldsymbol{h}^*}$ with~${\boldsymbol{g}^*}$. By the rotational symmetry of the problem,~${P(\boldsymbol{h} | \Delta)}$ depends only on~${|\boldsymbol{h}|}$.

The distribution of feasible regions for the latents~${\boldsymbol{x}_\mu}$ follows similarly. For a fixed~${\Delta}$, Eqs.~\eqref{eq:13} and~\eqref{eq:14} define the region~${X_\mu}$ of feasible latents~${\boldsymbol{x}_\mu}$ as a function of~${\boldsymbol{h}_{T(\mu)}}$ and~${\boldsymbol{h}_{R(\mu)}}$, so its distribution is that of~${X\left(\boldsymbol{h}^*,\boldsymbol{g}^*\right)}$ under the weights of Eq.~\eqref{eq:A11}.

Obtaining a distribution for the latents themselves requires us to specify how each~${\boldsymbol{x}_\mu}$ should be chosen from inside its feasible region. In simulations, we compute~${X_\mu}$ for the solution found by the optimizer, draw~${\boldsymbol{\hat{x}}_\mu}$ uniformly at random from the directions along which~${X_\mu}$ is nonempty, and then set~${|\boldsymbol{x}_\mu|}$ to be its minimal value along that direction. For consistency, the theoretical distribution of~${|\boldsymbol{x}_\mu|}$ is computed in the same manner by sampling Gaussian~${\left(\boldsymbol{z},\boldsymbol{y}\right)}$, computing~${X\left(\boldsymbol{h}^*,\boldsymbol{g}^*\right)}$, and weighting each sample by~${w}$ (Fig.~\ref{fig:8}C).

\subsection{The large-vocabulary limit: $\beta\gg1$}

The RS and 1-RSB cases proceed identically as follows. In the~${q_1\rightarrow0}$ limit, the constraint in~${I}$ is always active, since it is with high probability advantageous to make~$\left|\boldsymbol{h}\right|$ and~$\left|\boldsymbol{g}\right|$ as small as possible. The solution is piecewise based on whether~${\boldsymbol{h}}$ lies in the interior or on the surface of the unit ball. The resulting leading free energy is the same as for fixed~${\boldsymbol{\hat{x}}}$, Eqs.~(\ref{eq:A9}-\ref{eq:A14}), and hence the solution has the same leading large-${\beta}$ capacity,
\begin{align}
    & \alpha \approx C_p(\kappa)\frac{R}{\log\beta}.
\end{align}
Thus replica symmetry breaking changes the finite-${\beta}$ capacity quantitatively but not the leading logarithmic dependence on vocabulary size.

\subsection{The LoRA-only limit: $\Delta\rightarrow\infty$}

As before, we take~$\Delta\rightarrow\infty$, resulting in~$\kappa\rightarrow0$. In the RS case, the inner minimization is solved by taking~${h_1\rightarrow1}$, and the leading free energy reduces to
\begin{equation}
\begin{split}
g\left(q\right) &=-\frac{1}{2}Rq+\frac{1}{2}\alpha\left(1+Rq-2\sqrt{q}\left\langle\left|\boldsymbol{z}\right|\right\rangle_{\boldsymbol{z}}\right)\\
&+\frac{1}{2}\left(\beta-\alpha\right) \left\langle\left[\sqrt{q}\left|\boldsymbol{z}\right| -1\right]_+^2\right\rangle_{\boldsymbol{z}}
\end{split}
\end{equation}

In the 1-RSB case, the inner minimization is solved by taking~$\boldsymbol{h}\rightarrow\boldsymbol{\hat{z}}$. The solution is
\begin{align}
I\left(\boldsymbol{z},\boldsymbol{y},\Delta,q_1,\kappa\right) & \rightarrow\left(1-\sqrt{q_{1}}\left|\boldsymbol{z}\right|\right)^{2}+\left[\sqrt{q_{1}}\left|\boldsymbol{y}\right|-1\right]^{2}_{+}
\end{align}
The leading free energy becomes
\begin{equation}
\begin{split}
&g\left(q_{1},\gamma\right)=-\frac{R}{2}\gamma\log\left(1+q_{1}/\gamma\right)\\
&\quad-\alpha\gamma\log \left\langle \exp\left(-\frac{1}{2\gamma}\left(\sqrt{q_{1}}\left|\boldsymbol{z}\right|-1\right)^{2}\right)\right\rangle_{\boldsymbol{z}}\\
&\quad-\left(\beta-\alpha\right)\gamma\log\left\langle\exp\left(-\frac{1}{2\gamma}\left[\sqrt{q_{1}}\left|\boldsymbol{z}\right|-1\right]^{2}_{+}\right)\right\rangle_{\boldsymbol{z}}
\end{split}
\end{equation}

\begin{figure}[!t]
\centering{}\includegraphics[width=0.8\columnwidth]{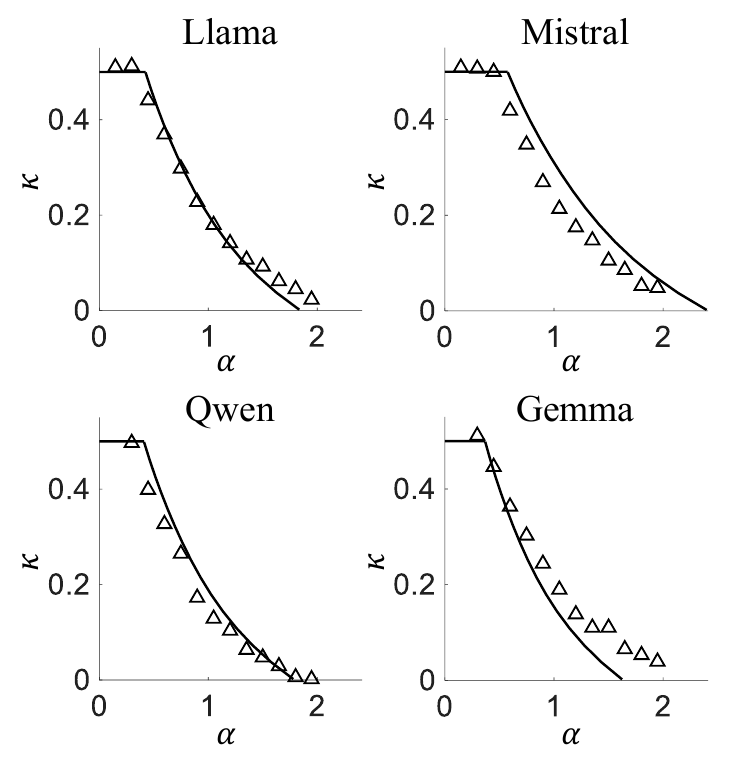}\caption{Margin~${\kappa}$ versus capacity~${\alpha}$ in simulations of the pair model using the readout layers of four LLMs in place of a Gaussian readout (triangles), compared with the 1-RSB theory for a Gaussian readout at each model's vocabulary size~${M}$ (lines). Capacity is varied through~${N}$ at fixed~${P=600}$, ${R=2}$, and~${p(\Delta)}$ is the uniform distribution on~${[0,3]}$.}\label{fig:13}
\end{figure}

The bound~${\alpha=R}$ arises from the~$\gamma\rightarrow0$ branch. On this branch, the capacity conditions have solutions for~${\beta-\alpha}$ not too large. For~${g}$ we obtain
\begin{equation}
g\left(q_{1},\gamma\right) = \frac{R-\alpha}{2}\gamma\log\gamma + \gamma g_0\left(q_{1}\right) + O(\gamma^2\log\gamma),
\end{equation}
where
\begin{equation}
\begin{split}
g_0\left(q_{1}\right) =&  \frac{R(R-1)}{2}\log(2\pi q_1) - R\log |S^{R-1}| + \frac{R}{2q_1}\\
&-\left(\beta-R\right)\log\left( 1 - H_R(1/\sqrt{q_1}) \right).
\end{split}
\end{equation}
Here,~${H_R(a) = \text{Pr}\left(|\boldsymbol{z}|>a\right)}$ is the upper radial tail probability for a standard ${R}$-dimensional Gaussian vector, and~${|S^{R-1}|}$ denotes the surface area of the unit sphere~${S^{R-1}}$. Imposing both~${g=0}$ and~${\partial_\gamma g = 0}$ and subtracting the two resulting equations yields
\begin{equation}
\label{eq:A12}
g_0(q_1) = 0 ,
\end{equation}
and
\begin{align}
    \alpha=R + O\left(\gamma\log\gamma\right) \rightarrow R .
\end{align}
This matches the result of the counting argument in Sec.~\ref{sec:LoRA-only}. Solutions to Eq.~\eqref{eq:A12} exist only for~${\beta}$ sufficiently close to~${\alpha}$, and make up the vertical~${\alpha=R}$ branches in Fig.~\ref{fig:4}C--D. Numerically, we find that the two branches~${\gamma=0}$ and~${\gamma>0}$ are mutually exclusive: for any~${\beta}$, a solution exists on exactly one branch.

\section{Real LLM readouts}\label{sec:appendix_D}

The theory in the preceding sections assumes that the LLM readout can be replaced by an i.i.d. Gaussian one. To test this assumption, we repeated the pair model simulations with the rows of the readout matrix~${A}$ taken from the readout layers of four pretrained LLMs (Llama-3.1-8B-Instruct, Mistral-7B-Instruct-v0.3, Qwen2.5-7B-Instruct, and Gemma-2-9B-it), keeping the synthetic pair biases. In each case, the simulations remain close to the Gaussian-readout theory evaluated at that model’s vocabulary size~${M}$ (Fig.~\ref{fig:13}).

\section{Effective rival biases inferred from the full model}\label{sec:appendix_E}

\begin{figure}[htbp]
\centering\includegraphics[width=\columnwidth]{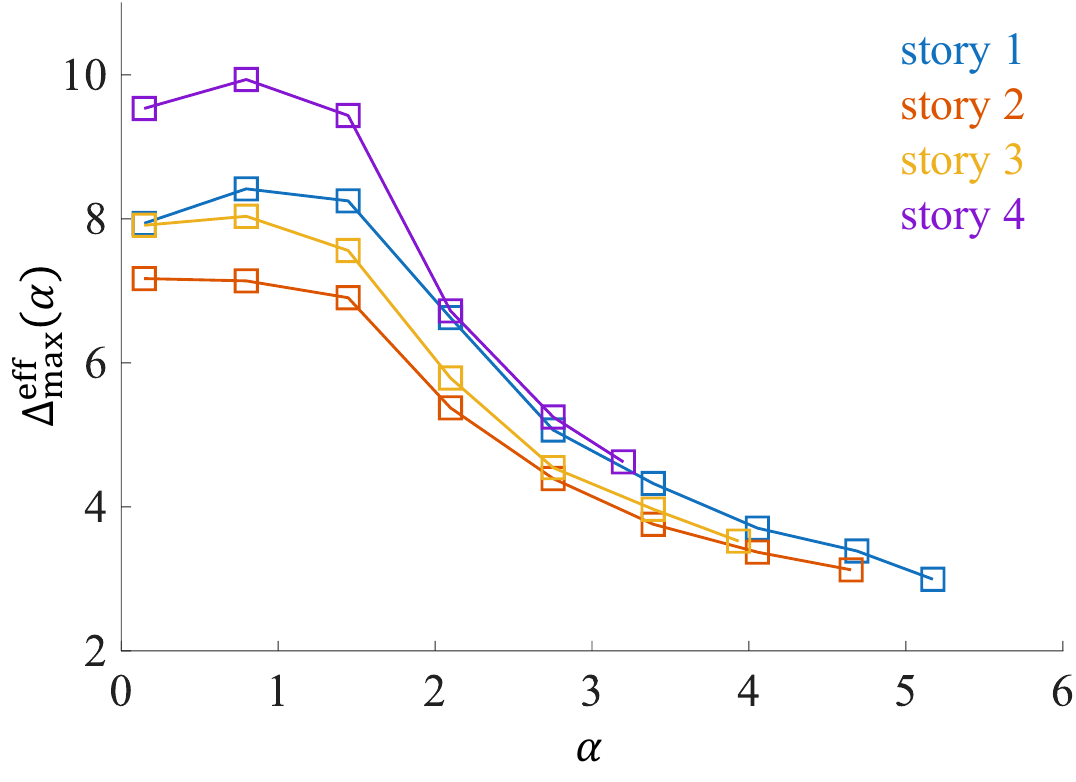}
\caption{Effective maximum rival bias~${\Delta_{\max}^{\mathrm{eff}}}$ inferred from the full LLM versus capacity~${\alpha}$. Each point is computed from the mean margin and capacity at the corresponding point of Fig.~\ref{fig:6}A, with~${P=600}$ and the full LLM vocabulary~${M=128{,}256}$. The pair-model bias distribution is taken to be uniform on~${[0,\Delta_{\max}]}$, with each story’s own~$\bar{n}$.}\label{fig:14}
\end{figure}

The pair model can also be used to interpret the LLM data. To make the comparison, we take~${p\left(\Delta\right)}$ uniform on~${[0,\Delta_{\max}]}$, and invert the capacity conditions at each point~${\left(\alpha,\beta,\kappa\right)}$ measured in the LLM (Fig.~\ref{fig:6}A), using each story's own~${\bar{n}}$, to obtain an effective maximum rival bias~${\Delta_{\max}^\mathrm{eff}}$ (Fig.~\ref{fig:14}). This effective bias tends to increase with~${N}$, which might suggest that the margin is limited more heavily by the most extreme negative biases at large~${N}$, and by the bulk of typical negative biases at small~${N}$.

\section{Numerical methods}\label{app:methods}

Simulations maximize the margin~${\kappa}$ jointly over~${U}$, ${X}$, and~${\kappa}$ with Adam in double precision, using a log-sum-exp relaxation of the storage constraints whose inverse temperature is annealed geometrically from~1 to~${10^6}$. Each step uses a batch of randomly sampled distractors, always including the rivals in the pair model. The normalization~${\max_n|\boldsymbol{h}_n|\le1}$ is enforced by a penalty~${\lambda\sum_n[|\boldsymbol{h}_n|^2-1]_+}$ and a final rescaling of~${U}$. A solution is accepted only if its worst margin over the full vocabulary is within~${10^{-6}}$ of~${\kappa}$; latent norms are then chosen within their feasible regions as described in Appendix~C. Capacities are found by bisection in~${N}$, using nested subsets of the feature neurons so that feasibility is monotonic in~${N}$, and the margin at each~${N}$ in the margin--capacity curves is the maximum reached in a separate full-length run. Optimization settings are listed in Table~\ref{tab:opt}. Points without error bars are single runs or means of two runs.

\begin{table}[b]
\caption{Optimization settings: Adam learning rate, penalty strength~$\lambda$, distractor batch size (``full'' means all distractors), and number of steps. In all cases, $U$, $X$, and~$\kappa$ are initialized with variances~$0.01$ and~$1$ and at~$0.1$, respectively, and Gaussian readout entries have variance~$1/4096$.}
\label{tab:opt}
\begin{ruledtabular}
\begin{tabular}{lcccc}
 & Rate & $\lambda$ & Batch & Steps \\
\hline
LoRA-only & $4\times10^{-4}$ & $10^{-6}$ & 2000 & $10^{6}$ \\
Pair, fixed~$\hat{\boldsymbol x}_\mu$ & $4\times10^{-4}$ & $0.1$ & full & $2\times10^{6}$ \\
Pair, trained~$\hat{\boldsymbol x}_\mu$ & $2\times10^{-4}$ & $5\times10^{-4}$ & $10^{4}$ & $2\times10^{6}$ \\
LLM & $10^{-3}$ & $0.1$ & $2000$ & $2\times10^{6}$ \\
\end{tabular}
\end{ruledtabular}
\end{table}

Stories are Wikipedia articles about events in 2024--2025, after the training cutoff of Llama-3.1-8B-Instruct, selected to be at least 601 tokens long, to exclude list-like pages, and to span topics including sports, natural disasters, politics, science, and culture. LLM simulations use four such articles truncated to~${P=600}$ tokens, and the analysis of Fig.~\ref{fig:9} uses~100 more. LoRA-only and pair-model simulations use Gaussian readouts, with synthetic biases in the pair model. Simulations with other LLM readouts (Appendix~\ref{sec:appendix_D}) use their unnormalized readout matrices, with hidden dimensions 4096 (Llama, Mistral) and 3584 (Qwen, Gemma).

\bibliography{refs}

\end{document}